\documentclass[preprint2]{aastex}
\usepackage{epsfig}

\shortauthors{Chang et al.}
\shorttitle{The Young Pulsar J1357-6429 and Its Pulsar Wind Nebula}

\begin{document}

\title{X-ray Observations of the Young Pulsar J1357--6429 and Its Pulsar Wind Nebula}

\author{Chulhoon Chang\altaffilmark{1}, George G.\ Pavlov\altaffilmark{1,2}, 
Oleg Kargaltsev\altaffilmark{3}, and Yurii A.\ Shibanov\altaffilmark{2,4}}
\altaffiltext{1}{Department of Astronomy \& Astrophysics, Pennsylvania State University, PA 16802, USA; chchang@astro.psu.edu, pavlov@astro.psu.edu }
\altaffiltext{2}{St.-Petersburg State Polytechnical University, Polytekhnicheskaya ul., 29, 195251, Russia}
\altaffiltext{3}{Department of Astronomy, University of Florida, FL 32611, USA;
oyk100@astro.ufl.edu}
\altaffiltext{4}{Ioffe Physico-Technical Institute, Politekhnicheskaya 26, St.-Petersburg, 194021 Russia; shib@astro.ioffe.rssi.ru}

\begin{abstract}
We observed the young pulsar J1357--6429 with the {\it Chandra} and {\it XMM-Newton} observatories. The pulsar spectrum fits well a combination of absorbed power-law model ($\Gamma=1.7\pm0.6$) and blackbody model ($kT=140^{+60}_{-40}$ eV, $R\sim2$ km at the distance of 2.5 kpc). Strong pulsations with pulsed fraction of $42\%\pm5\%$, apparently associated with the thermal component, were detected in 0.3--1.1 keV. Surprisingly, pulsed fraction at higher energies, 1.1--10 keV, appears to be smaller, $23\%\pm4\%$.  The small emitting area  of the  thermal component either corresponds to a hotter fraction of the neutron star (NS) surface or indicates inapplicability of the simplistic blackbody description. The X-ray images also reveal a pulsar-wind nebula (PWN) with complex, asymmetric morphology comprised of a brighter, compact PWN surrounded by the fainter, much more extended PWN whose spectral slopes are $\Gamma=1.3\pm0.3$ and $\Gamma=1.7\pm0.2$, respectively. The extended PWN with the observed flux of $\sim7.5\times10^{-13}$ erg s$^{-1}$ cm$^{-2}$ is a factor of 10 more luminous then the compact PWN. The pulsar and its PWN are located close to the center of the extended TeV source HESS J1356--645, which strongly suggests that the VHE emission is powered by electrons injected by the pulsar long ago. The X-ray to TeV flux ratio, $\sim0.1$, is similar to those of other relic PWNe. We found no other viable candidates to power the TeV source.  A region of diffuse radio emission, offset from the pulsar toward the center of the TeV source, could be synchrotron emission from the same relic PWN rather than from the supernova remnant. 
  
\end{abstract}

%%%%%%%%%%%%%%%%%%%%%%%%%%%%%
%%%%%%%%%%%%%%%%%%%%%%%%%%%%%

\section{Introduction}

 Thanks to their rich observational manifestations, young energetic pulsars are among the most attractive targets for the {\sl Chandra} and {\sl XMM-Newton} observatories. These manifestations include thermal emission from the neutron star (NS) surface, non-thermal emission from the pulsar magnetosphere, and a pulsar-wind nebula (PWN) produced by the interaction of the pulsar wind with the ambient medium. The magnetospheric and thermal emission can differ in their relative strengths and exhibit distinct spectra and pulse profiles. The magnetospheric emission usually has power-law (PL) spectrum and is strongly pulsed. Detailed studies of several bright young pulsars indicate a nonuniform NS surface temperature distribution with cool ($T_C \lesssim 100$ eV) and hot ($T_H \sim 0.1-0.3$ keV)  thermal components, which are believed to be emitted from the bulk of the NS surface and polar caps, respectively \citep[e.g.][]{Pavlov2001,Zavlin2004,deLuca2005,Karga2005}. Since the sample of young pulsars with high quality X-ray spectra is still very small, further observations of young pulsars must be carried out to learn about the early epoch of NS cooling and test magnetospheric emission models.
  
In addition to X-ray emission from the NS surface and magnetosphere, a significant amounts of non-thermal emission is radiated from the PWN. PWNe display various shapes and structures, including (but not limited to) jets and tori, or cometary shaped tails \citep{Gaensler2006}. Observations with high angular resolution are particularly useful since they resolve the extended PWN emission, seen downstream of the termination shock, from the magnetospheric non-thermal emission.  Comparing properties of these two components, one can learn about the physical processes responsible for an efficient but yet unknown particle acceleration mechanism energizing PWNe. 
   
Although more than $\sim$70 PWNe have been observed with {\it Chandra} and {\it XMM-Newton} to date \citep[][hereafter KP08 and KP10]{Karga2008,Karga2010}, the existing models \citep{Kennel1984,Komissarov2004,Swaluw2005,Bucciantini2005} still cannot explain some of the observed complex morphologies. Pulsars of similar ages and spin-down luminosities can have PWNe with quite different properties. As of now,  it is not clear whether these differences can be solely attributed to different properties of the surrounding interstellar medium (ISM) or different intrinsic properties of pulsars also play an important role. A larger sample of well-resolved X-ray PWNe associated with pulsars having different intrinsic properties and residing in different environments  is needed to further advance our understanding of pulsar winds.
   
The young pulsar PSR J1357--6429 (hereafter J1357; $P$ = 166 ms, $\dot{E}$ = 3.1 $\times$ 10$^{36}$ erg s$^{-1}$, surface magnetic field $B$ = 7.8 $\times$ 10$^{12}$ G) was discovered in the Parkes Multibeam Pulsar Survey by \citet{Camilo2004}. The spin-down age, $\tau \equiv P/(2 \dot{P})$ = 7.3 kyr, indicates that J1357 is one of the youngest pulsars known. The distance of 2.5 kpc, estimated from  the pulsar's dispersion measure (DM = 127.2 cm$^{-3}$ pc) and the Galactic free electron density model \citep{Cordes2002}, implies a spin--down flux $\dot{E}/(4 \pi d^2)$ = 1.2 $\times$ 10$^{-9}$ erg s$^{-1}$ cm$^{-2}$.  

The pulsar is located within the TeV source HESS J1356--645 \citep[hereafter HESS J1356;][]{Renaud2008,Abramowski2011}.
The extended TeV emission is detected up to $12'$ from the peak of the TeV surface brightness, located at  R.A.$ =13^{\rm h}56^{\rm m}$, decl.$=-64^{\circ}30'$. Given the relatively small angular separation, $\sim 7'$, between J1357 and the center of the HESS source, and the TeV luminosity, $L_{\rm TeV}\sim 6 \times 10^{33}$ ergs s$^{-1} \sim 0.002\dot{E}$ (at the 2.5 kpc distance), it seems plausible that the HESS source could be powered by this pulsar. The TeV spectrum of HESS J1356 fits PL with the photon index $\Gamma_{\rm TeV}\approx2.2$,
without a cutoff at high energies. A possible detection of J1357 with {\sl AGILE}  at lower $\gamma$-ray energies (100 MeV -- 10 GeV) was reported by \citet{Pellizzoni2009}\footnote{After our paper was submitted, \citet{Lemoine2011} reported results of observations of J1357 with {\sl Fermi} LAT. They detected $\gamma$-ray pulsations and measured the photon index $\Gamma_{\rm GeV}\approx 1.5$, found an exponential cut-off at $\approx 0.8$ GeV, and estimated the luminosity $L_{\rm GeV} \approx 2\times 10^{34}$ erg s$^{-1}$, for $E>0.1$ GeV.}.
No optical counterpart to the pulsar or its PWN has been detected yet \citep{Mignani2011}.

At radio frequencies, \citet{Duncan1997} reported the extended emission from  the SNR candidate G309.8--2.6 , which is spatially coincident with HESS J1356. The extent of the radio emission appears to be smaller than that of the TeV source.
 J1357 has been previously observed with {\sl XMM-Newton} (2005 August 5; 15 ks; PI F.\ Camilo) 
and {\sl Chandra} HRC-S (2005 November 18 and 19; 16 and 17 ks; PI M.\ Mendez, observer L.\ Kuiper).
The analysis of those data was reported by \citet{Zavlin2007} and \citet{Esposito2007}\footnote{It should be noted that first report on the {\sl Chandra} results was published in ``Chandra News'' (March 2007, Issue number 14, pages 12-13 (see \url{http://cxc.harvard.edu/newsletters/news\_14/newsletter14.html}) by R.\ Kraft and A.\ Kenter. In particular, evidence for the presence of extended emission (PWN) and the detection of pulsed signal (pulsed fraction $40\%\pm 12\%$) were reported in that publication.}. These authors found that the pulsar spectrum likely consists of thermal and nonthermal components. \citet{Zavlin2007} found pulsations
with the radio pulsar period and pulsed fraction of $63\%\pm 15\%$, while \citet{Esposito2007} reported only an upper limit of 30\% on modulation amplitude. 
\citet{Zavlin2007} also reported a tail-like extended emission in the HRC-S image,
with a luminosity of $2.5\times 10^{31}$ erg s$^{-1}$ (for $d=2.5$ kpc), while
\citet{Esposito2007} concluded that the surface brightness distribution around the pulsar ``is consistent with that from a point source'' and estimated a 2--10 keV upper limit of $3\times 10^{31}$ erg s$^{-1}$ on the PWN luminosity.
To resolve these controversies, reliably disentangle various emission components, and study the relation between the pulsar and HESS~J1356, we have carried out deeper observations with the {\sl Chandra} ACIS and {\sl XMM-Newton} EPIC detectors.

The details of the observations and data analysis are presented in Section 2. We discuss  possible interpretations of the PWN morphology, describe inferences from the pulsar spectrum,  and speculate on the nature of HESS~J1356 and its relation to the other sources in the field in Section 3.

\section{Observations and Data Analysis}
We observed J1357 for 60 ks with the ACIS detector on board {\sl Chandra} on 2009 October 8 (ObsID 10880). The target was imaged on the I3 chip, $\approx 30''$ from the aim point. The other chips operated during the observation were I0, I1, I2, S2, and S3. The observation was carried out in timed exposure mode and telemetered in Very Faint format. The detector was operated in full frame mode, which provides time resolution 3.24 s. The useful effective exposure time (live time) is 59.22 ks. There were no significant particle background flares during the observation. The data were reduced and analyzed with the {\sl Chandra} Interactive Analysis Observations (CIAO) package (ver.4.2), with CALDB 4.1.4. We used {\sl Chandra} Ray Tracer (ChaRT)\footnote{See http://cxc.harvard.edu/chart/threads/index.html.} and MARX software\footnote{See http://space.mit.edu/CXC/MARX/.} for image analysis and XSPEC (ver.12.5.0) for spectral analysis.

%Fig.1
\begin{figure}[h]
\centering
\includegraphics[scale=0.3]{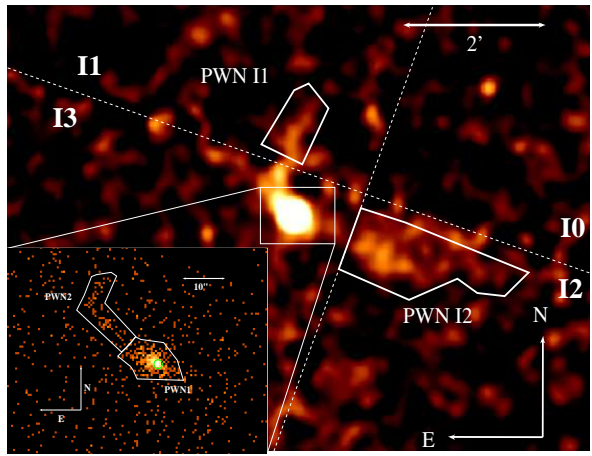}
\caption{{\sl Chandra} images of J1357 and its PWN in the 0.3--8.0 keV band. The exposure-corrected larger image is binned to a pixel size of 0$\farcs$98 and then smoothed with the Gaussian kernel of about 3$''$ radius using the {\it aconvolve} script. The dashed lines show the chip boundaries. PWN\,I1 and PWN\,I2 are the extraction regions (see text). The zoomed unsmoothed image in the inset shows the extraction regions for the image and spectral analysis.}
\label{fig1}
\end{figure}

J1357 was also observed with the {\sl XMM-Newton} EPIC MOS1, MOS2 and PN detectors on 2009 August 14 and 15 (ObsID 0603280101). The two MOS detectors were operated in Full Window mode with a time resolution 2.6 s, while the PN detector was in Small Window mode, which provides 6 ms time resolution (at the expense of
reduced efficiency, $\approx 71\%$).
The observation was processed with the {\sl XMM-Newton} Science Analysis Software (SAS; ver.9.0.0). The total live times of observations were 78.1 ks and 55.5 ks for EPIC MOS and PN, respectively. 
However, the data were contaminated by strong flares, which  exceeded the quiescent count rate (0.1 counts/s for MOS and 0.2 counts/s for PN) by a factor of up to 8. After excluding contaminated data, the useful scientific exposure times are 62.2 and 44.6 ks for MOS and PN, respectively. The medium filter was used for all the EPIC detectors. The data were filtered to allow only standard event grades (patterns $\leq$ 12 for MOS and $\leq$ 4 for PN).

\subsection{Images}

\subsubsection{Chandra}

To produce {\it Chandra} images of the pulsar and its vicinity at subpixel resolution, we removed the pipeline pixel randomization and applied the subpixelization tool which improves quality of the images using grades of split-pixel events \citep{Mori2001,Tsunemi2001}. Figure~\ref{fig1} shows the {\sl Chandra} image of the region around J1357 and the extraction regions on the ACIS-I3 chip used for the image and spectral analysis in the 0.3--8 keV energy band. In the image we see a bright pointlike source surrounded by diffuse emission. The position of the center of the brightest pixel is R.A. = 13$^{\rm h}$57$^{\rm m}$02$\fs$496, decl.=$-64\arcdeg29\arcmin30\farcs$06 (J2000.0). The radio pulsar position is R.A.=13$^{\rm h}$57$^{\rm m}$02$\fs$43(2), decl.=$-64^\circ29'30\farcs2(1)$ \citep{Camilo2004}. The difference of 0$\farcs$45 between the X-ray and radio positions is smaller than the error in absolute {\it Chandra} astrometry (0$\farcs$6 at the 90\% confidence level\footnote{See http://cxc.harvard.edu/cal/ASPECT/celmon/}).  We measure the background on the I3 chip in the 50$''$ radius circle centered at about 1$\farcm$5 southeast from the source. The bright pointlike source contains 371 counts in the $r=1''$ aperture, including 12\% contribution from the background and the diffuse emission around the source. 

The diffuse emission in the immediate vicinity of the pulsar (region PWN1; shown in the inset in Figure~\ref{fig1}), excluding the region of  $r=1\farcs5$ circle centered on the pulsar, has 282 total counts in 91 arcsec$^{2}$, including 4.6\% background contribution. The background--subtracted surface brightness within the PWN1 region is $\simeq$ $2.97\pm0.19$ counts arcsec$^{-2}$. We also define PWN2 region (see the same inset in Figure~\ref{fig1}), which includes the diffuse feature extending northeast of the pulsar and turning northward  at $\sim23''$ from the pulsar. This region contains 74 counts in 108 arcsec$^{2}$, including 21\% background contribution. This corresponds to the background-subtracted surface brightness of $\simeq$ $0.54\pm0.08$ counts arcsec$^{-2}$. 

%Fig.2
\begin{figure}[h!]
\centering
\includegraphics[scale=0.32,angle=90]{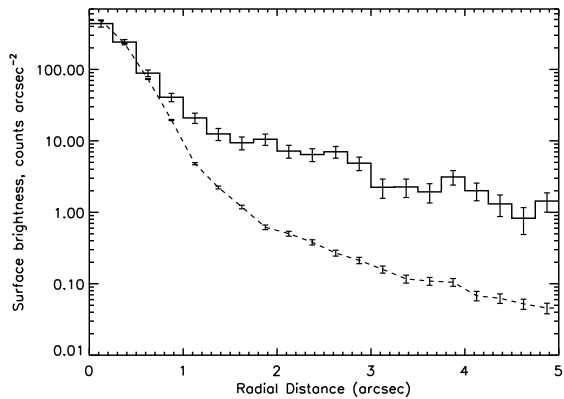}
\caption{Radial profiles of the observed emission (solid histogram with bin size 0$\farcs$25) around the pulsar and the simulated PSF (dashed line) in the 0.3--8.0 keV energy band. The PWN emission dominates the pulsar PSF at $r>0\farcs75$ from the pulsar.}
\label{fig2}
\end{figure}

The larger exposure-corrected, binned and smoothed image (Figure~\ref{fig1}) shows an even fainter diffuse emission extending $\sim$ 2$\farcm$5 westward and over 1$\farcm$5 northward. The PWN region on I1 chip (PWN\,I1; $\approx$1921 arcsec$^2$) includes 295 total counts, of which about 70\% comes from the background, estimated from the 150$''$ radius circle at about 5$'$ north-northeast of the source region. The total number of counts in the PWN region on I2 chip (PWN\,I2; $\approx$6128 arcsec$^2$) is 1036, including 72.5\% contribution from the background, estimated from the 100$''$ radius circle located at just south of the source region. The background-subtracted PWN surface brightness within the source regions on I1 and I2 chips is $0.046\pm0.009$ and $0.047\pm0.006$ counts arcsec$^{-2}$, respectively.

The radial profile of the emission centered on the brightest pixel position is plotted in Figure~\ref{fig2}, with the simulated point spread function (PSF) shown  for comparison. To simulate the PSF, we used the ChaRT and MARX packages following the standard procedure. To reduce the statistical errors of the simulation, the normalization of the input spectrum was increased by a factor of 100 while running ChaRT. The simulated images were then rescaled back for direct comparison with the data. We have also tried several values of the MARX Dither Blur parameter  and found an optimal value of 0$\farcs$25, which minimizes the difference  between the simulated PSF and the observed count distribution in the immediate vicinity of the pulsar position. The histogram in Figure~\ref{fig2} shows that the PWN starts dominating  the unresolved pulsar emission at $\ga$ 0$\farcs$75 from the pulsar.

%Fig.3
\begin{figure}[h!]
\centering
\includegraphics[scale=0.4]{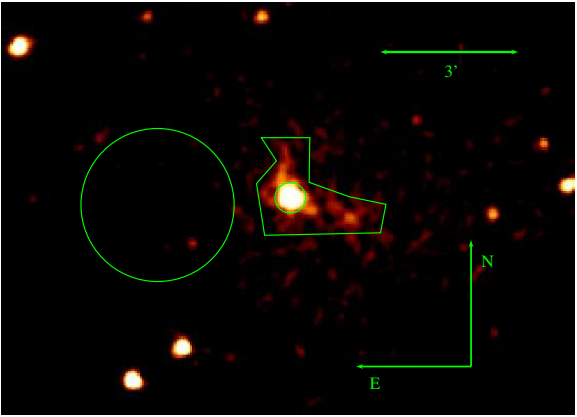}
\caption{{\it XMM-Newton} EPIC MOS1 + MOS2 image of J1357 and its PWN in the 0.3--10 keV energy band. The image is binned to a pixel size of 3$''$ and then smoothed with a Gaussian kernel of 9$''$ radius. Extraction regions ($r=20''$ circle for the pulsar region, 11210 arcsec$^2$ polygon for the PWN region, and $r=100''$ circle for the background region) for the spectral analysis are also shown.}
\label{fig3}
\end{figure}

\subsubsection{XMM-Newton}
\begin{figure}[h!]
\centering
\includegraphics[scale=0.4]{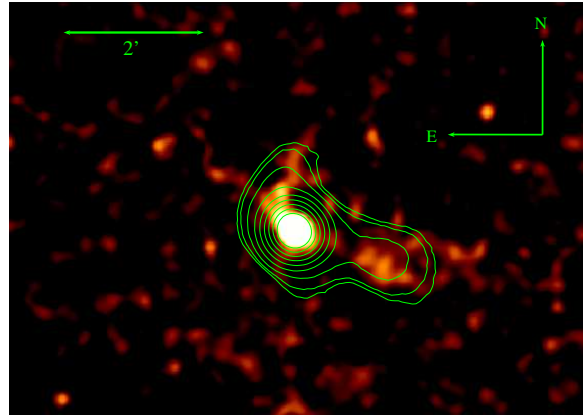}
\caption{Smoothed {\it Chandra} ACIS image from Figure~\ref{fig1} with contours of {\it XMM-Newton} MOS1+2 image overlaid. The contours correspond to the brightness levels of 2.9, 3, 3.3, 4, 5, 6, 7, 8, and 9 counts.}
\label{fig4}
\end{figure}

\begin{figure}[h!]
\centering
\includegraphics[scale=0.4]{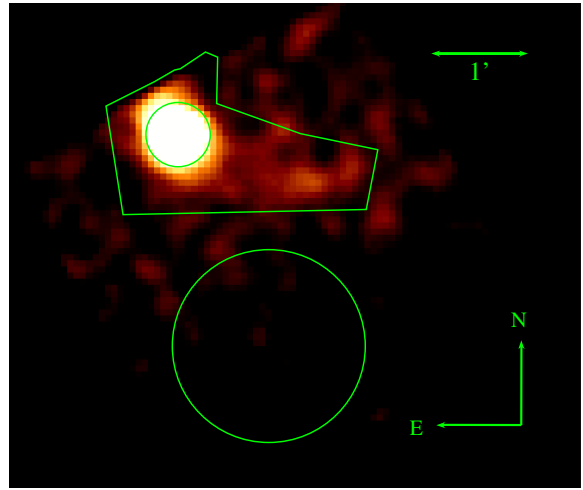}
\caption{{\it XMM-Newton} EPIC PN image is binned to a pixel size of 3$''$ and then smoothed with a Gaussian kernel of about $9''$ radius. The image also shows the $r=20''$ circle for the pulsar region, 9159 arcsec$^2$ polygon for the PWN region, and $r=60''$ circle for the background region for the spectral analysis of J1357 and its PWN in 0.3--10 keV. Unfortunately, the northern part of the PWN emission cannot be seen because the source is near the chip edge.}
\label{fig5}
\end{figure}

 Figure~\ref{fig3} shows the combined  EPIC  MOS1+MOS2 image (in 0.3--10 keV) and several regions that we use in our analysis. Within the $r=20''$ circle centered on the pulsar  we find 1,900 MOS1+MOS2  counts, of which the background (taken from the $r= 100''$ circle east of the pulsar) contributes about $15\%$. The MOS images show the large-scale PWN morphology reminiscent of that seen in the {\it Chandra} image (cf.\ Figure~\ref{fig1}). Indeed, the contours from the MOS1+2 PWN image match the {\it Chandra} PWN image quite well (see Figure~\ref{fig4}). Since the source is near the chip edge in the PN image, only part of the extended emission can be seen, while the other part is out of the field of view (see Figure~\ref{fig5}). We choose an $r= 20''$ circle centered on the pulsar in the PN image for the source region, and a 60$''$ radius circle, centered at about $2\farcm5$ south-southwest from the source for the background region. The total number of counts in the source region is 2,403, including 19\% background contribution.
 
%Fig.4

%Fig.5

To choose an optimal PWN region in the {\it XMM-Newton} MOS and PN images, we simulated the PSF using Quicksim{\footnote{See http://heasarc.gsfc.nasa.gov/docs/xmm/quicksim/quicksim.html.}}. The simulation shows that an $r=30''$ circle around the position of the pulsar contains 88\% encircled count fraction for the point source. Excluding this circle from the PWN polygon region (see Figure 3) leaves 3,446  counts in MOS1+2, including  $\sim60$\% background contribution. Although some regions of the PWN are out of the PN Small Window field of view (see  Figure~\ref{fig5}), we still collected  3,771 counts, including $\sim70$\% background contribution. 

%Table.1
\begin{table*}
\footnotesize
\center
\caption{Spectral fits of the field sources in the {\it Chandra} and {\it XMM-Newton} images\label{tab1}}
\vspace{0.5cm}
\setlength{\tabcolsep}{1mm}
\begin{tabular}{ccccccccc}\hline\hline
Source&$n_{\rm H,21}$&$\Gamma$&$\mathcal{N}$\tablenotemark{a}&$kT$\tablenotemark{b}&$\mathcal{N}_{T}$\tablenotemark{c}&$F_{\rm abs}$\tablenotemark{d} &$F_{\rm unabs}$\tablenotemark{e} &($C$ or $\chi^2_{\nu}$)/dof\tablenotemark{f} \\ \hline
1&$12\pm4$&1.78$^{+0.39}_{-0.35}$&$4\pm5$&$\cdots$&$\cdots$&$14\pm1$&$28^{+12}_{-6}$&1.05/30 \\
2&3$^{+4}_{-3}$&$\cdots$&$\cdots$&1.6$^{+0.5}_{-0.3}$&2$^{+2}_{-1}$&$9\pm2$&$10\pm2$&0.88/17 \\
3&3$^{+4}_{-3}$&$\cdots$&$\cdots$&$1.1\pm2$&6$^{+7}_{-3}$&$6\pm1$&$6\pm1$&0.92/16 \\
4&3$^{+6}_{-3}$&$\cdots$&$\cdots$&1.0$^{+0.3}_{-0.2}$&3$^{+5}_{-2}$&3.3$^{+0.8}_{-0.7}$&3.7$^{+0.9}_{-0.7}$&9.16/8 \\
5&3$^{+4}_{-3}$&1.9$^{+0.8}_{-0.6}$&9$^{+19}_{-9}$&$\cdots$&$\cdots$&$3.0^{+0.9}_{-0.8}$&$5^{+4}_{-1}$&0.98/11 \\
6&$7\pm2$&1.54$^{+0.27}_{-0.22}$&$40\pm10$&$\cdots$&$\cdots$&$17\pm2$&$25^{+5}_{-3}$&0.84/39 \\
\hline\hline
\end{tabular}
\tablecomments{The errors shown represent 90\% confidence intervals.}\\ 
\tablenotetext{\rm a}{~PL normalization in units of 10$^{-6}$ photons cm$^{-2}$ s$^{-1}$ keV$^{-1}$ at 1 keV.}
\tablenotetext{\rm b}{~BB temperature in keV.}
\tablenotetext{\rm c}{~BB normalization $\mathcal{N}_{T}=10^{-3}R^2_{\rm km}/D^2_{10}$, where $R_{\rm km}$ is the source radius in km, and $D_{10}$ is the distance to the source in units of 10 kpc.}
\tablenotetext{\rm d}{~Absorbed flux in the 0.3--8 keV, in units of $10^{-14}\, \rm erg\, cm^{-2} \, s^{-1}$.}
\tablenotetext{\rm e}{~Unabsorbed flux in the 0.3--8 keV, in units of $10^{-14} \, \rm erg\, cm^{-2} \, s^{-1}$.}
\tablenotetext{\rm f}{~Best-fit $C$-statistic value (source 4) or reduced $\chi^2$ value (other sources).}
\end{table*}

\subsubsection{Other sources in the field of Chandra and XMM-Newton}
\begin{figure*}[htp]
\begin{center}
\includegraphics[width=65mm]{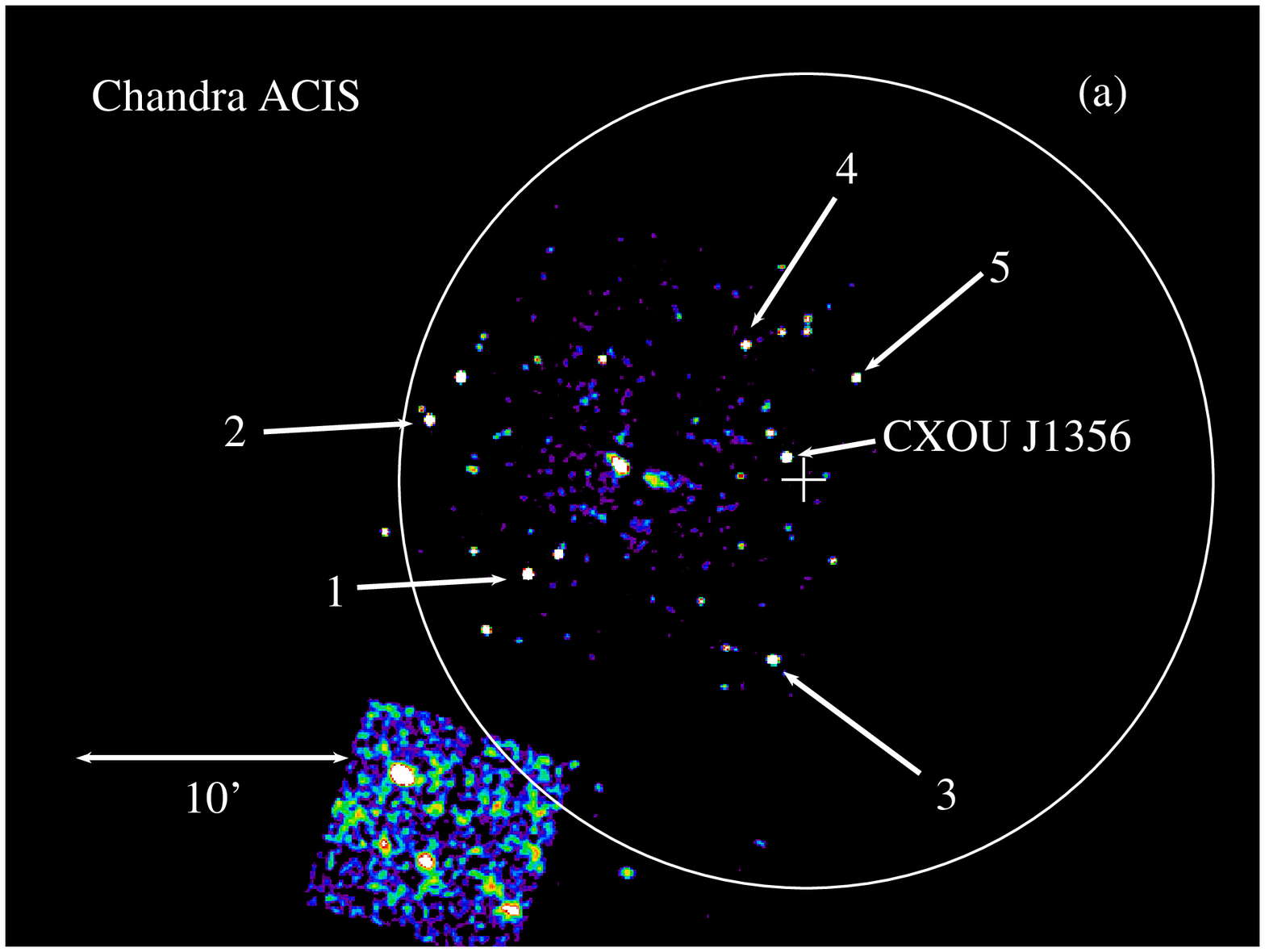}
\includegraphics[width=65mm]{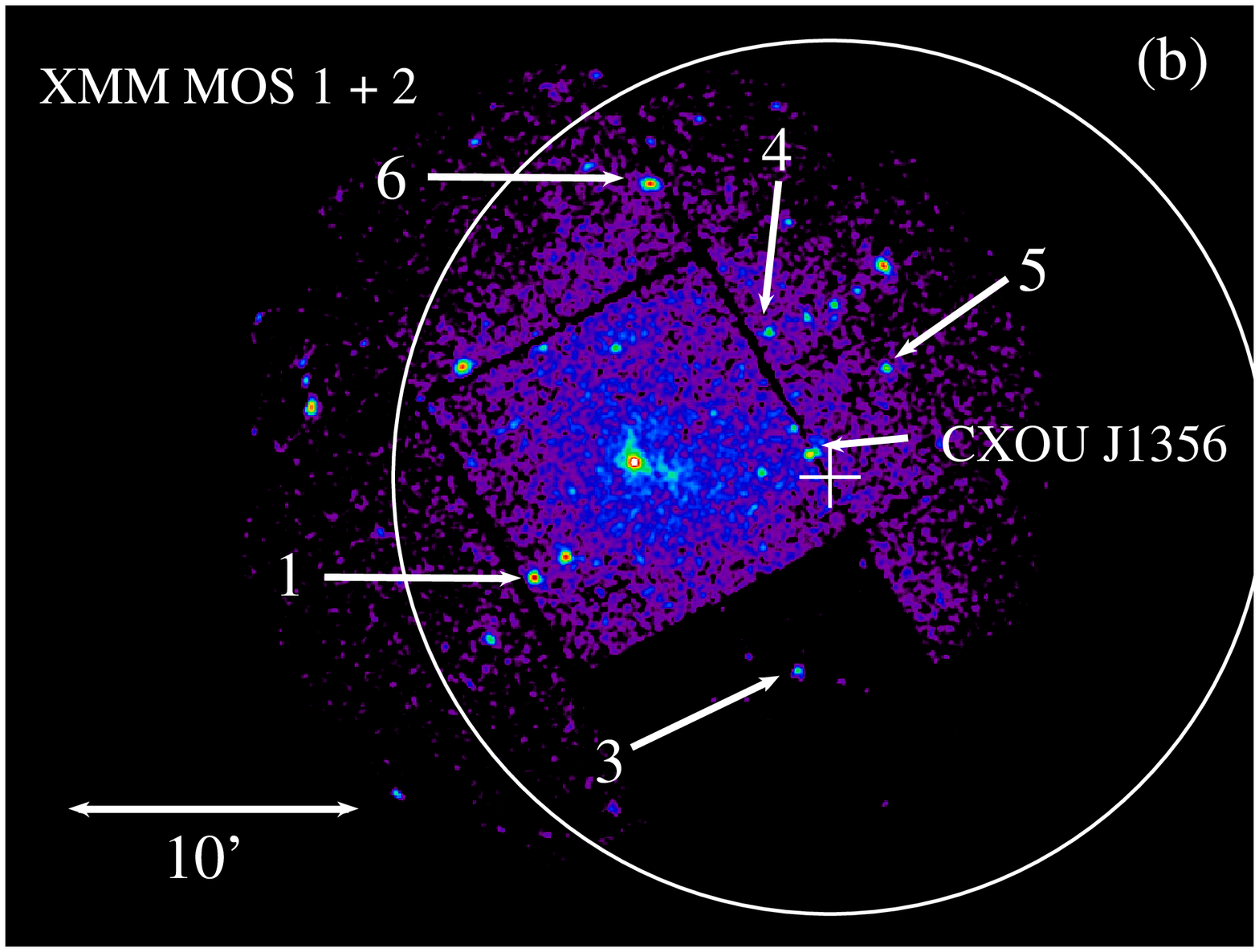}
\includegraphics[width=65mm]{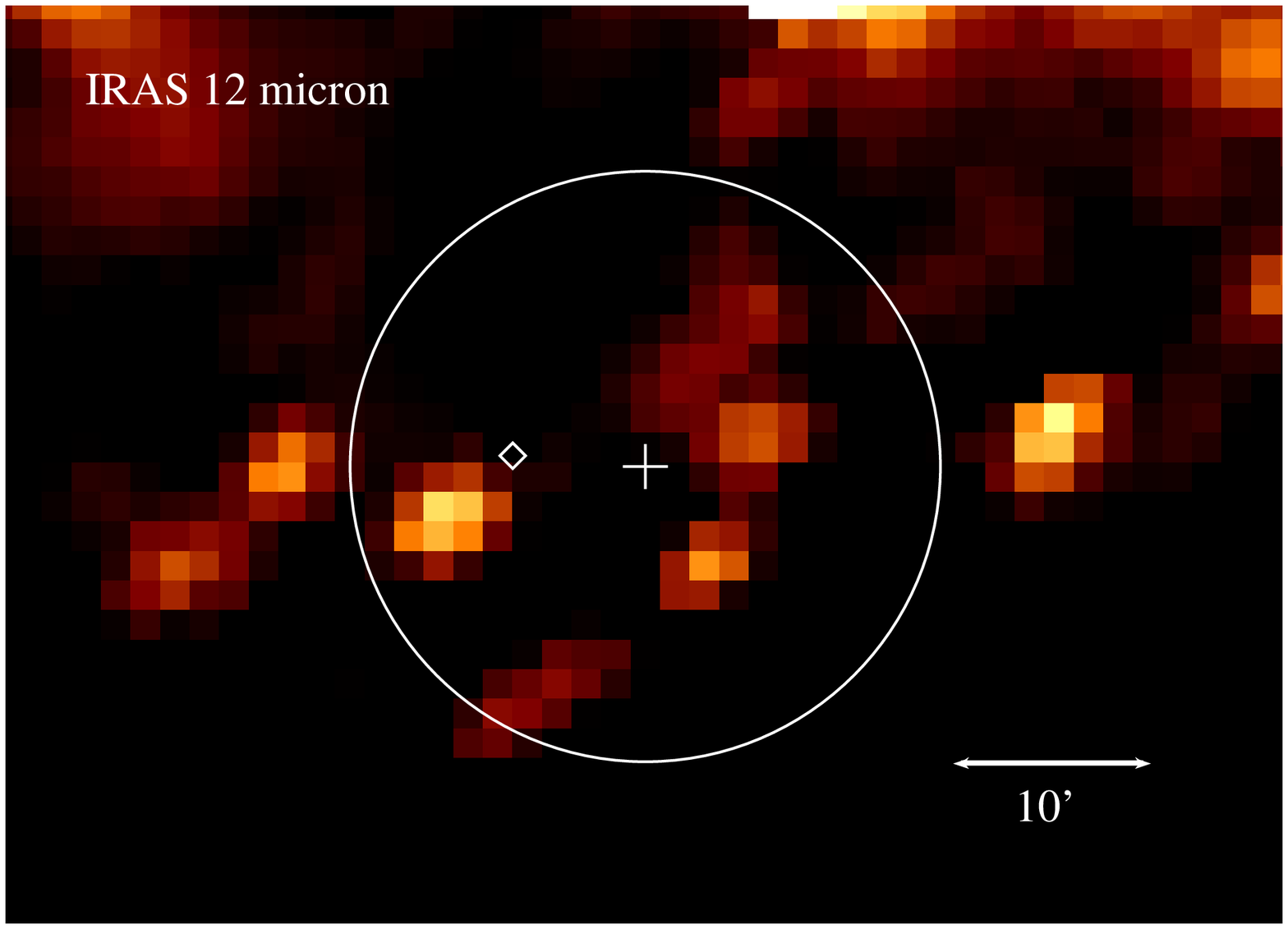}
\includegraphics[width=65mm]{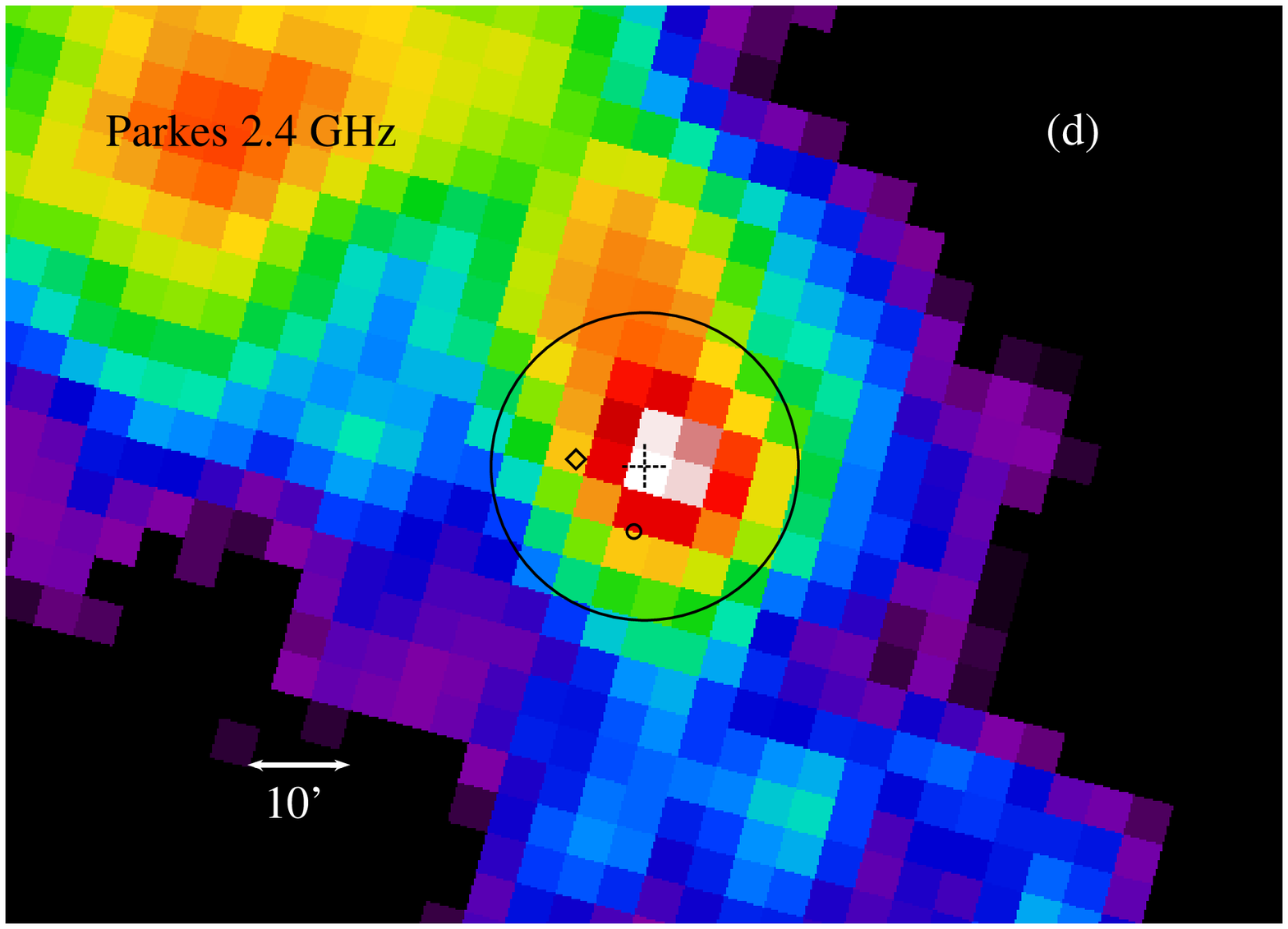}
\includegraphics[width=65mm]{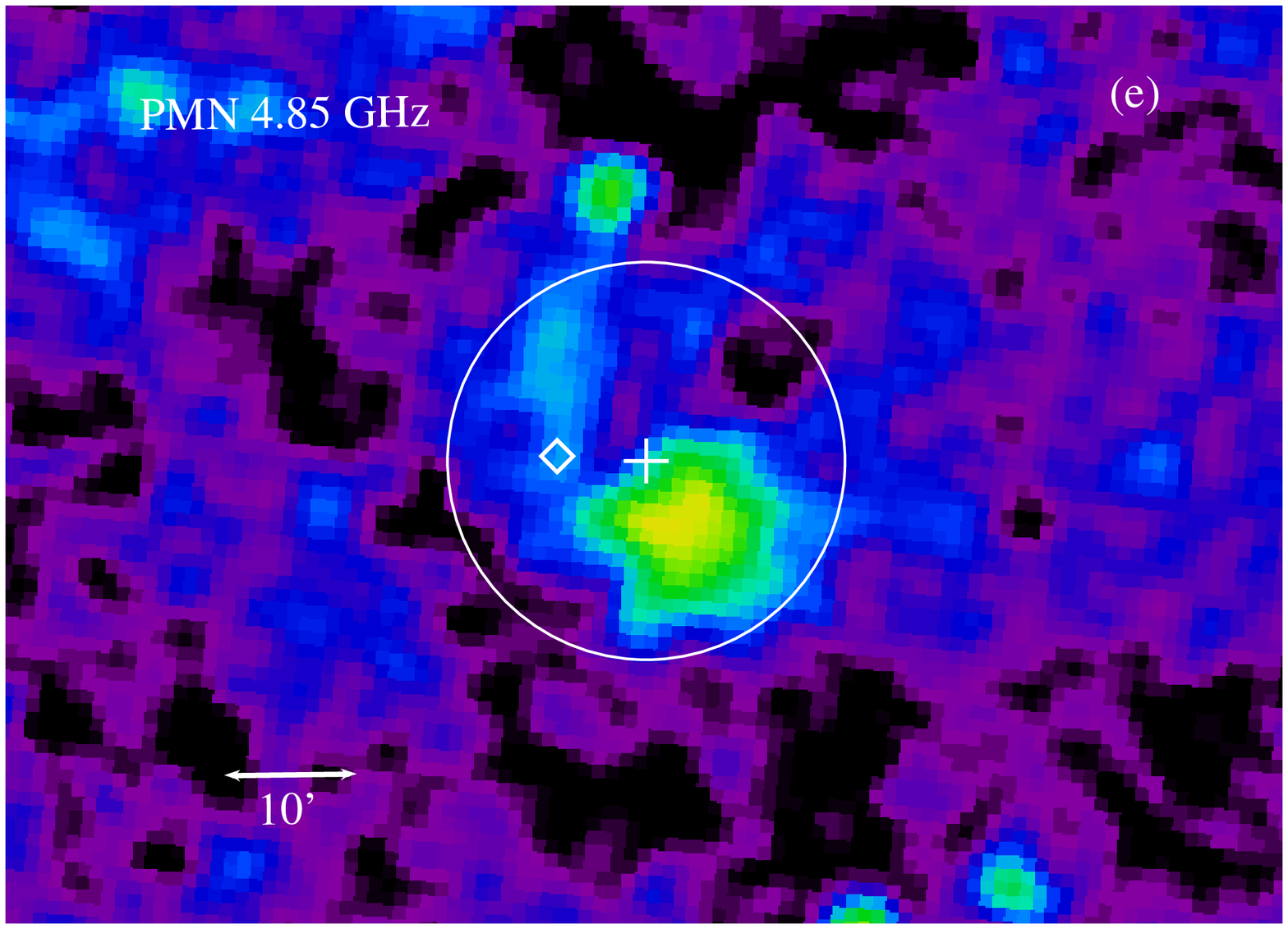}
\includegraphics[width=65mm]{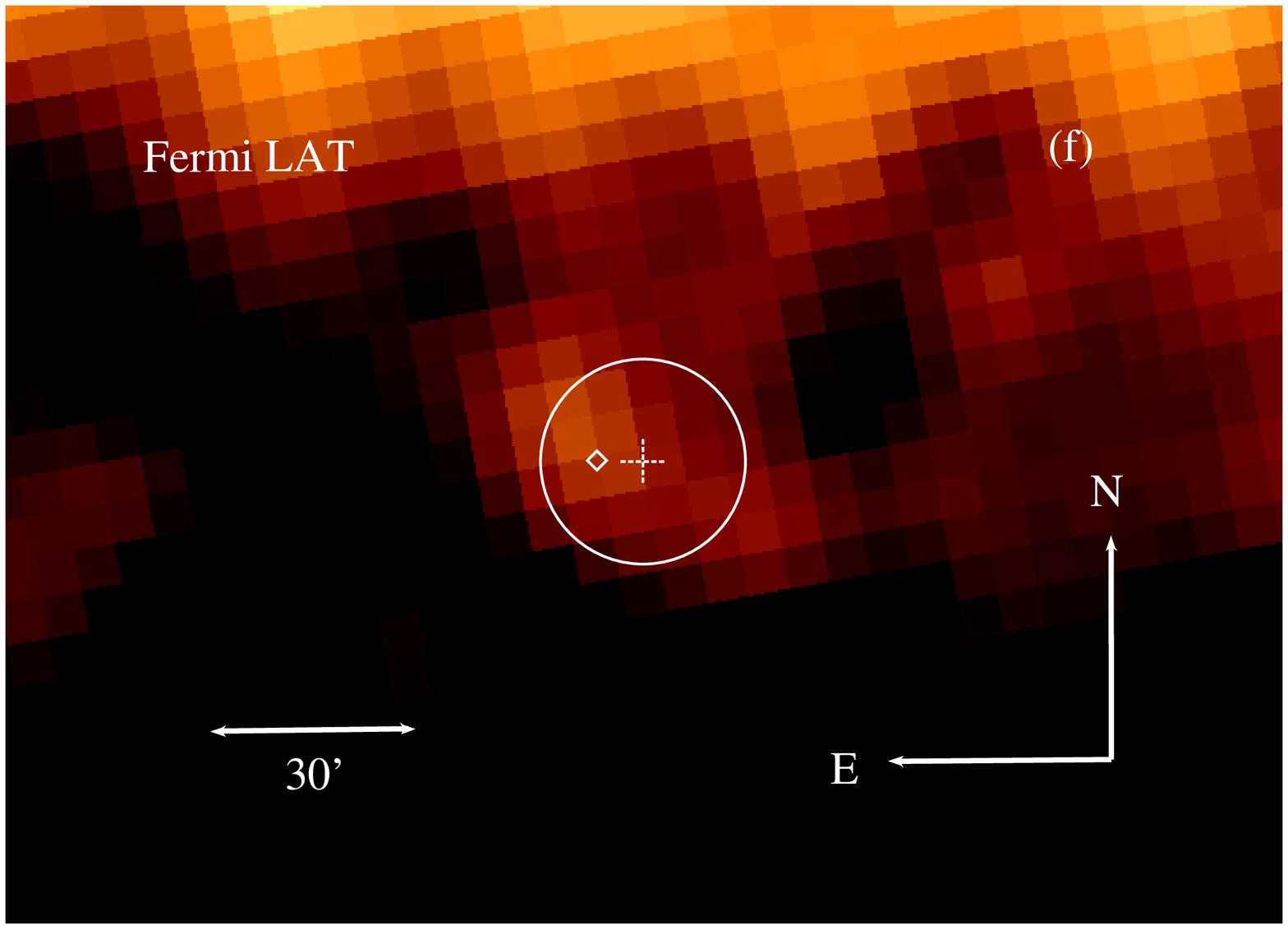}
\end{center}
\caption{X-ray, IR, and radio images of the J1357 and HESS J1356 field. The cross shows the centroid position of HESS J1356, while the circle (15$'$ radius) indicates the intrinsic (rms) Gaussian width of the source, according to \cite{Renaud2008}. The white arrows in panels (a) and (b) indicate CXOU J135605.9--642909 and 6 other sources (see text for details). (a) {\sl Chandra} ACIS image in 0.5--8 keV. (b) XMM MOS1+2 image in the 0.3--10 keV. (c) 12 micron IRAS image. (d) 2.4 GHz Parkes image. The small circle located south of the cross point indicates the position of SNR candidate G309.8-2.6. (e) 4.85 GHz PMN image. (f) Fermi LAT image. The diamond in panels c--f shows the position of J1357.} 
\label{fig6}
\end{figure*}

Panels a and b of Figure~\ref{fig6} show that J1357 is the brightest extended X-ray source within the imaged part of the HESS J1356 field. Nevertheless, we have examined the other sources in the ACIS FOV. Three brightest sources on S3 chip southwest of the circle in Figure~\ref{fig6}a have Two Micron All Sky Survey \citep[2MASS;][]{Skrutskie2006} counterparts. Most likely, they are stars unrelated to the TeV source. There are five relatively bright sources inside the circle, which do not have 2MASS counterparts (Sources 1--5 in Figure~\ref{fig6}a). Four of these sources are seen in both the ACIS and MOS images, while source 2 fell into the chip gap in MOS. One more source, lacking a 2MASS counterpart, is seen in the MOS images (Source 6 in Figure~\ref{fig6}b), but it happens to be outside the  ACIS field of view.  We fitted an absorbed PL model to the spectrum of each of these sources. When the PL model was unacceptable, we fitted the spectrum with the blackbody (BB) model. The spectral properties of the sources are summarized in Table~\ref{tab1}.  These sources could be CVs or quiescent LMXBs, or they could be AGNs seen through the Galactic disk.

Finally, we investigated in detail the relatively bright source on ACIS I0 chip, CXOU J135605.9--642909 (hereafter CXOU J1356), which is apparently the same source as 1RXS J135605.5--642902. The source is also seen in the EPIC images, but it is near the chip gap in the EPIC MOS1 and MOS2 images. We extracted 443 total counts within the 5$''$ radius around the brightest pixel of the source from the ACIS image. The spectrum was binned to have minimum 25 counts per bin. We did not obtain an adequate fit with a single component (PL or BB) model (e.g., $n_{\rm H} \approx 2.0\times10^{21}$ cm$^{-2}$, $\Gamma \approx 3.4$, and $\chi^2_{\nu}=1.8$ for 14 degrees of freedom (dof) for the PL model). The spectrum fits better the two-component PL+BB model, which gives $n_{\rm H} \approx 4.3\times10^{21}$ cm$^{-2}$, $\Gamma = 2.9$, $kT = 0.12$ keV and $\chi^2_{\nu}=1.1$ for 12 dof. The absorbed flux is $6.4\times10^{-14}$ erg cm$^{-2}$ s$^{-1}$ in the 0.3--8 keV band. We found no significant variability in the light curve. We have searched the field at other wavelengths to understand the nature of the source. The nearest optical/IR source was found in the Naval Observatory Merged Astrometric Dataset \citep[NOMAD;][]{Zacharias2005} and 2MASS catalogs ($B=13.81$, $V=13.66$, $R=12.69$, $J=12.199\pm0.029$, $H=11.779\pm0.031$, and $K=11.768\pm0.030$), offset by $\approx 0\farcs4$ from the position of CXOU J1356. The optical-NIR colors show that the source might be a G or K star. The estimated X-ray to optical flux ratio is about $4\times10^{-3}$, which is similar to a typical value ($\sim 10^{-3}$) for a G or K star \citep{Maccacaro1988}. 

Although many more sources are seen in the ACIS and MOS images, the low numbers of counts preclude spectral fitting, making it even more difficult to establish the nature of these sources. None of the investigated sources shows evidence for extended emission in X-rays, and there are no compelling reasons to believe that any of them is related to HESS J1356.
 
%Fig.6

\subsubsection{Multiwavelenths images}

In addition to analyzing the {\it Chandra} and {\it XMM-Newton} images, we have searched the archival radio and IR data covering the HESS~J1356 region. The 12 micron image (Fig.~\ref{fig6}c) from Infrared Astronomical Satellite (IRAS) shows several bright sources within the circle. Figure~\ref{fig6}d shows the 2.4 GHz image from the Parkes survey of the southern Galactic plane{\footnote{see http://www.atnf.csiro.au/research/surveys/2.4Gh\_Southern/data.html.}}. The extended radio emission has been previously classified as an SNR candidate G309.8--2.6 \citep{Duncan1997}. Another radio image from the Parkes-MIT-NRAO \citep[PMN{\footnote{see http://www.parkes.atnf.csiro.au/observing/databases/pmn/pmn.html.}};][]{Griffith1993} survey at 4.85 GHz also shows an extended emission southeast of HESS J1356 (see Fig.~\ref{fig6}e). The Fermi LAT image (0.5--100 GeV), shown in Figure 6f, reveals a faint GeV emission near the pulsar position. The angular extent of the GeV excess is consistent with that expected for the unresolved point source. No counterparts to J1357 or its X-ray PWN are apparent in the IR and radio images. Deeper observations of this region with higher spatial resolution would be desirable to understand  the nature of the extended radio and IR emission.

\subsection{Spectral and Timing Analysis}
\subsubsection{Pulsar Spectrum}

\begin{figure}[h!]
\centering
\includegraphics[scale=0.32,angle=90]{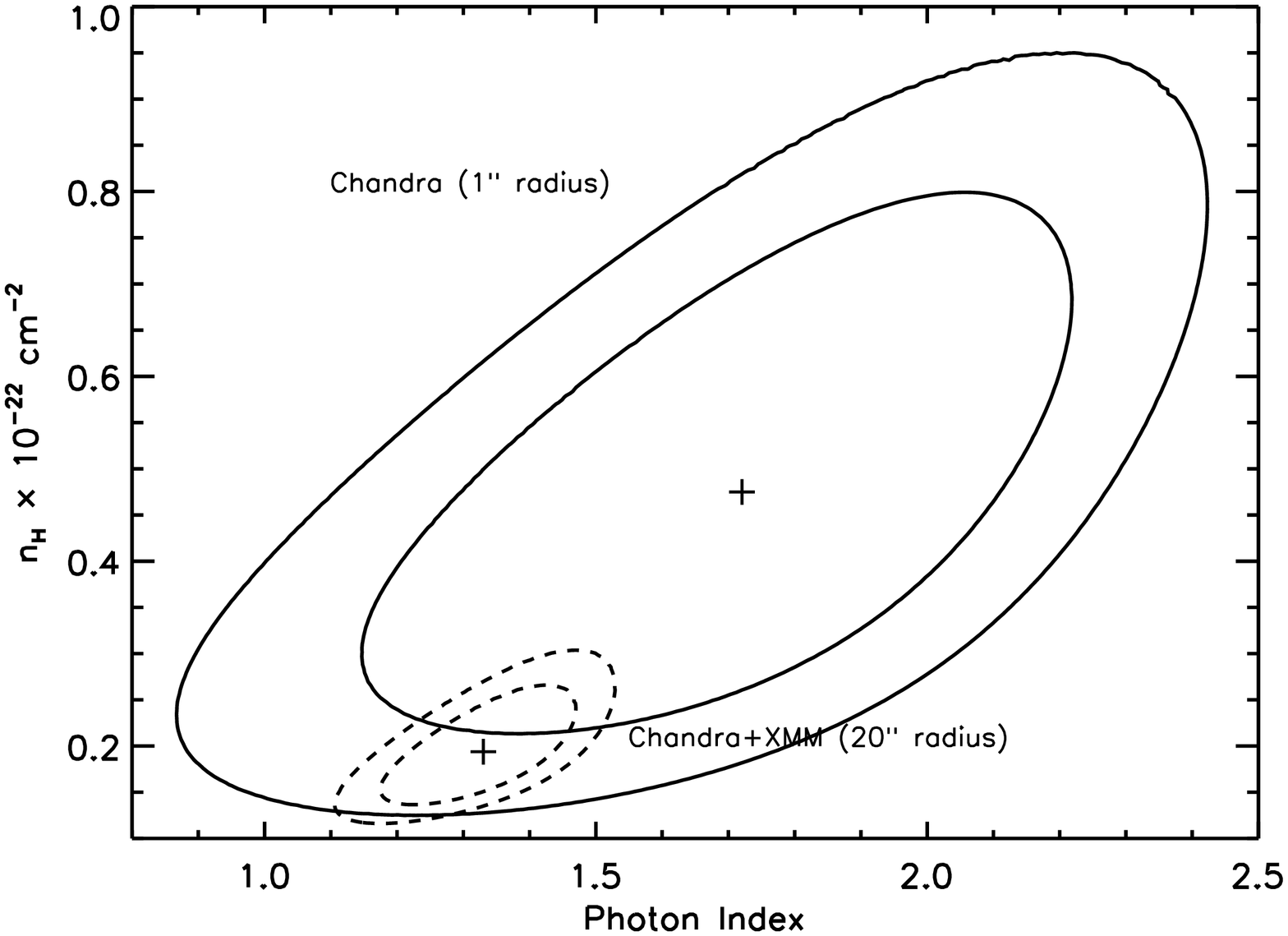}
\includegraphics[scale=0.32,angle=90]{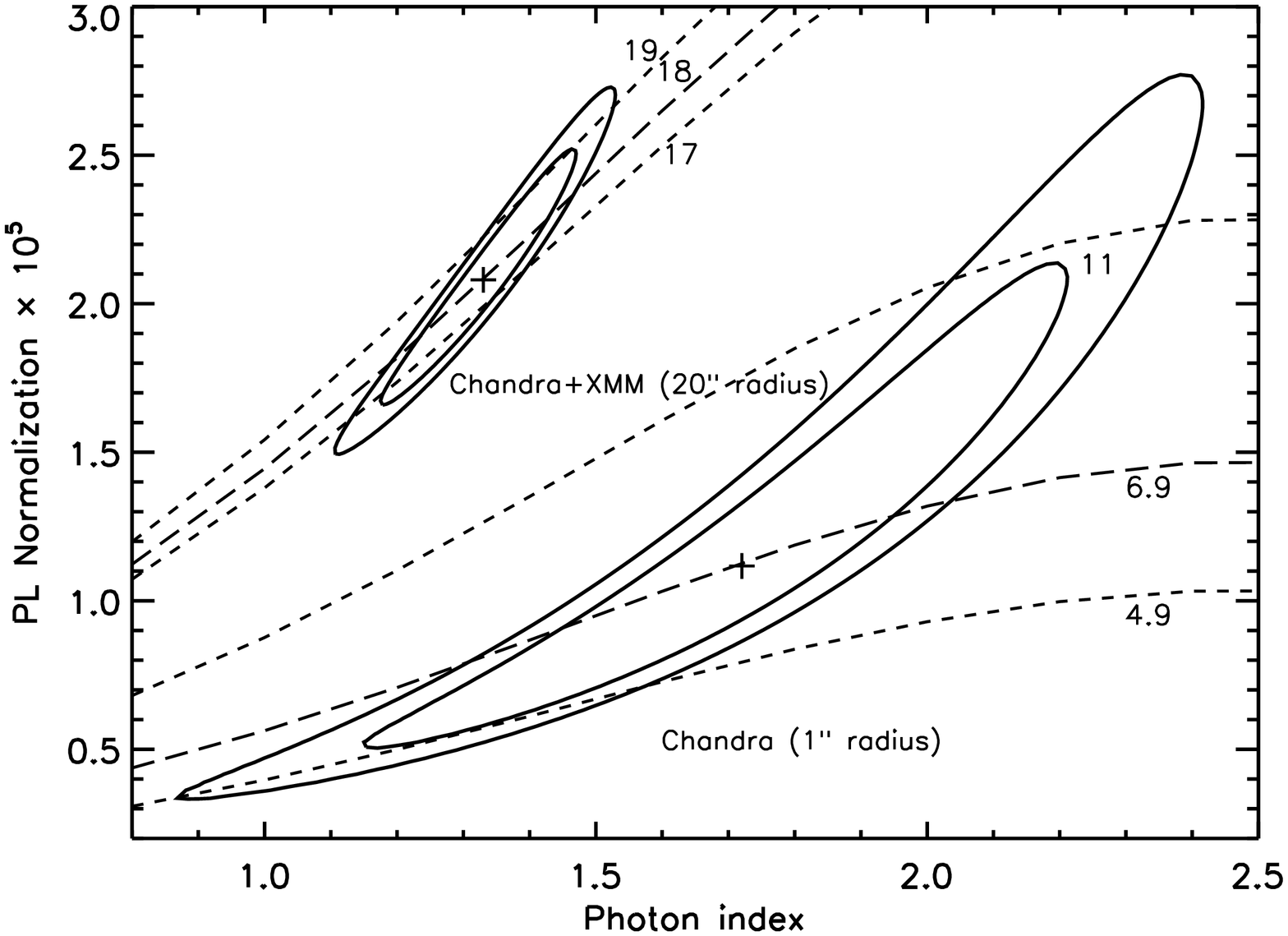}
\caption{{\it Top}: Confidence contours (68\% and 90\%) in the $\Gamma-n_{\rm H}$ plane for the PL+BB fit to the pulsar's spectrum from the {\sl Chandra} observation (1$''$ extraction radius) and {\sl Chandra} + {\it XMM-Newton} observations (20$''$ extraction radius). {\it Bottom}: Confidence contours (68\% and 90\%) for the PL component of the PL + BB fit to pulsar's spectrum from the {\sl Chandra} observation (1$''$ extraction radius) and {\sl Chandra} + {\it XMM-Newton} observations (20$''$ extraction radius). The PL normalization is in the units of 10$^{-5}$ photons cm$^{-2}$ s$^{-1}$ keV$^{-1}$ at 1 keV. The dashed curves are the lines of constant unabsorbed flux for the PL component in the units of 10$^{-14}$ ergs cm$^{-2}$ s$^{-1}$ in 0.5--8 keV.}
\label{fig7}
\end{figure}

\begin{figure}[h!]
\centering
\includegraphics[scale=0.32,angle=90]{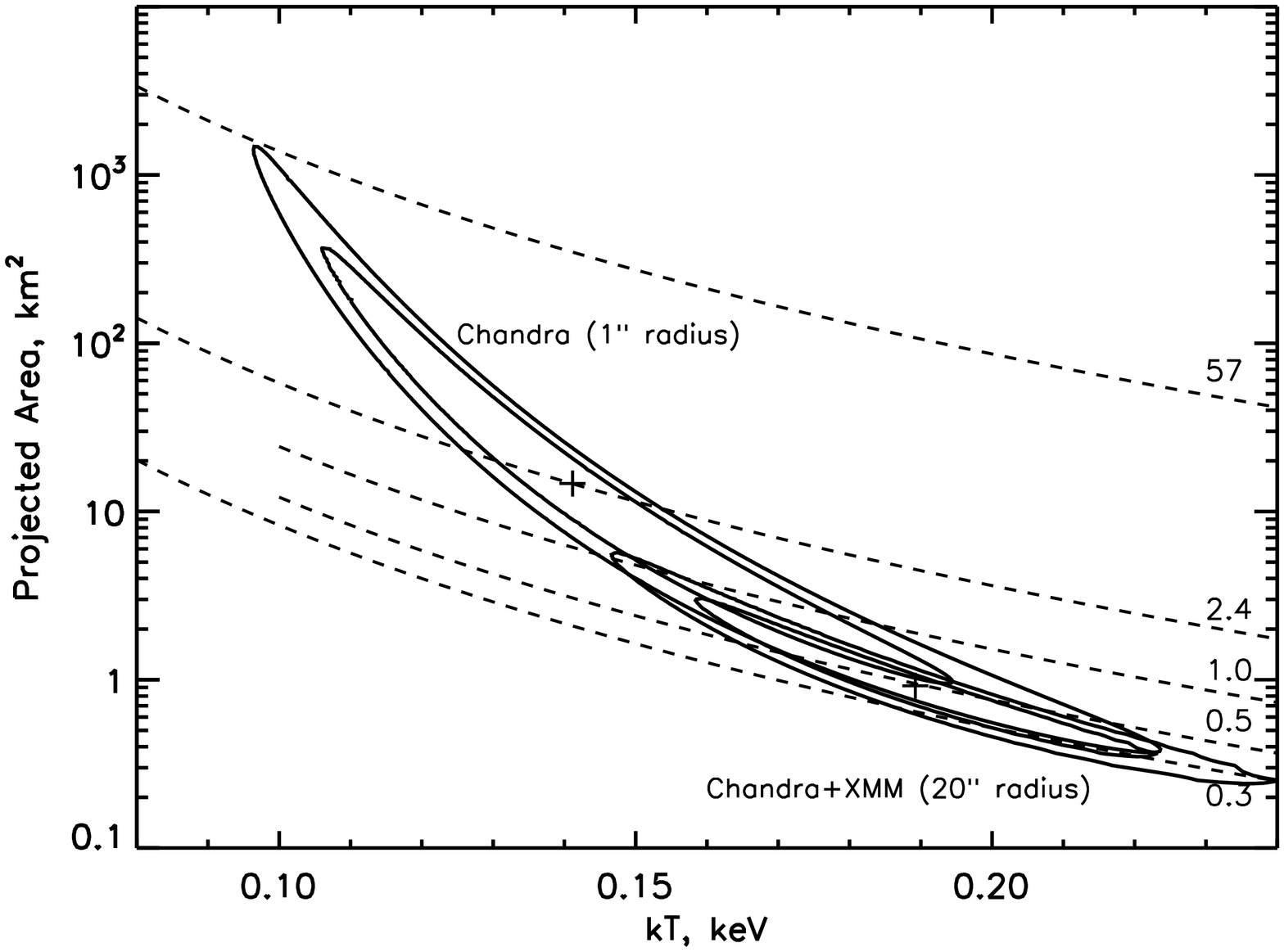}
\caption{Confidence contours (68\% and 90\%) for the BB parameters of the PL + BB fit to the pulsar's spectrum from the {\sl Chandra} (1$''$ extraction radius) observation and {\sl Chandra} + {\it XMM-Newton} observations (20$''$ extraction radius). The BB normalization is the projected emitting area (km$^{2}$). The dashed lines show the constant bolometric luminosity in units of 10$^{32}$ ergs s$^{-1}$, assuming $d=2.5$ kpc.}
\label{fig8}
\end{figure}

For the spectral analysis of the {\it Chandra} observation, we first extracted the pulsar's spectrum from a small circle (1$''$ radius aperture with 88\% encircled energy fraction) to reduce the contamination from the PWN (see Figure~\ref{fig1}). There are 371 total counts in this aperture, of which about 11\% come from the PWN and 1\% from the background. 
We binned the spectrum with minimum of 10 counts per bin and used the $C$ statistic \citep{Cash1979} in spectral fitting. 
The spectrum was first fitted with a single component model, such as the absorbed PL model and the absorbed BB model, in 0.3--8 keV. The PL fit gave $n_{\rm H,21}\tbond n_{\rm H}/(10^{21}$ cm$^{-2}) \approx 1.4$, and $\Gamma \approx 2.4$ ($C$ = 52 for 34 bins, 31 dof ), while the BB was not statistically acceptable ($C$ = 173 for 34 bins, 31 dof ). Since the PL fit quality was not good, we fit the same spectrum with the absorbed PL+BB model. The much better fit ($C= 28$ for 34 bins for 29 dof ) yields $n_{\rm H, 21} \approx 4.7$, $\Gamma \approx 1.7$, $kT \approx 0.14$ keV ($T \approx 1.6 \times 10^6$ K), and the projected emitting area $\mathcal{A} \approx 13$ $d^{2}_{2.5}$ km$^2$ ($d_{2.5}$ = d/2.5 kpc).
This temperature and area correspond to the apparent radius $R = \sqrt{\mathcal{A}/\pi} \sim 2d_{2.5}$ km and bolometric luminosity $L_{\rm bol} = 4\mathcal{A}\sigma T^4 \sim 2 \times 10^{32} d^{2}_{2.5}$ erg s$^{-1}$. The pulsar's observed flux is $F^{\rm abs}_{\rm psr} = 6.7^{+0.9}_{-0.8} \times  10^{-14}$ erg cm$^{-2}$s$^{-1}$ in 0.3--8 keV (corrected for the 88\% encircled energy fraction), while the corresponding unabsorbed flux is $F^{\rm unabs}_{\rm psr} = 32^{+190}_{-21} \times 10^{-14}$ erg cm$^{-2}$s$^{-1}$ in the same energy band. The unabsorbed luminosity is $L_{\rm psr}=2.4 \times 10^{32}d^{2}_{2.5}$ erg s$^{-1}$. The luminosity of the non-thermal (PL) component is $L_{\rm PL}=5.2^{+2.9}_{-1.5}\times10^{31}$ erg s$^{-1}$ ($\sim$ 22\% of the total unabsorbed luminosity), while the luminosity of the thermal (BB) component is $L_{\rm BB}=1.9^{+14}_{-1.4}\times10^{32}$ erg s$^{-1}$. The uncertainties of the fits are reported in Table~\ref{tab2}, and the confidence contours are shown in Figures~\ref{fig7} and \ref{fig8}.

%Fig.7

%Fig.8

Instead of BB, one can use a neutron star atmosphere (NSA) model for
the thermal component, which gives a lower (effective) temperature and
a larger size of the emitting region \citep{Pavlov1995}. Since the spectrum emerging from a magnetized atmosphere is anisotropic, one should know the geometry
of the surface magnetic field (e.g., the angles between the rotation axis, magnetic axis and line of sight for a dipole field) to fit the spectrum. For a crude estimate, however, one can use the NSA models in XSPEC, which are calculated for emergent flux assuming the magnetic field perpendicular to the surface. 
We made such an estimate using the hydrogen NSA models for the NS mass
$M=1.4 M_\odot$, radius $R=10$ km, and magnetic field $B=1\times 10^{13}$ G. If the distance is
fixed at $d=2.5$ kpc, the fit ($C=29$ for 30 dof) yields
$n_{\rm H,21} = 4.6^{+0.8}_{-0.7}$, $\Gamma = 1.45^{+0.50}_{-0.60}$,
$T_{\rm eff} = 0.96^{+0.04}_{-0.05}$ MK ($T_{\rm eff}^\infty = 0.74^{+0.03}_{-0.04}$ MK for the gravitationally redshifted temperature)\footnote{A similar fit was obtained by \citet{Lemoine2011}}. It shows that the thermal component can be emitted from a substantial part of the NS surface.

To compare the {\it Chandra}  ACIS spectrum with those obtained from the {\it XMM-Newton} EPIC detector, we extracted the pulsar spectrum from the 20$''$ radius aperture for MOS and PN and the background spectrum from the 100$''$ radius for MOS and 60$''$ radius for PN (see Figures~\ref{fig3} and \ref{fig5}). The spectrum is heavily contaminated by the compact PWN (about 42\% contribution). The spectra were binned with minimum 50 counts per bin for each detector. Since the absorbed PL fit was not satisfactory ($\chi^2_{\nu}$ = 1.94 for 82 dof), we fitted the absorbed PL+BB model which yields $n_{\rm H, 21} \approx 2.2$, $\Gamma \approx 1.35$, $kT \approx 0.18$ keV, and the projected emitting area $\mathcal{A} \approx 2 d^{2}_{2.5}$ km$^2$. The BB parameters correspond to the apparent radius $R \sim 0.8d_{2.5}$ km and $L_{\rm bol} = 8.6\times10^{31} d^2_{2.5}$ erg s$^{-1}$. Although the best-fit $kT$ and $\mathcal{A}$ from the {\it XMM-Newton} observations are different from those observed from the {\it Chandra} data, these parameters are consistent within their uncertainties. The observed flux (corrected for the 80\% encircled counts fraction) is $F^{\rm abs}_{\rm pulsar+PWN}= 2.6 \times 10^{-13}$ erg cm$^{-2}$s$^{-1}$, and the corresponding unabsorbed flux is $F^{\rm unabs}_{\rm pulsar+PWN} = 3.6 \times 10^{-13}$ erg cm$^{-2}$s$^{-1}$ in 0.3--10 keV. Additional details are provided in Table~\ref{tab2}.

%Table.2
\begin{table*}
\footnotesize
\center
\caption{Absorbed BB+PL fits to the pulsar (or pulsar + compact PWN) spectrum  from the {\it Chandra} and {\it XMM-Newton} observations. \label{tab2}}
\vspace{0.5cm}
\setlength{\tabcolsep}{1mm}
\begin{tabular}{ccccccccc}\hline\hline
Model\tablenotemark{a}&$n_{\rm H,21}$&$\Gamma$&$\mathcal{N}$\tablenotemark{b}&$kT$\tablenotemark{c}&$\mathcal{A}$\tablenotemark{d}&$F_{\rm abs}$\tablenotemark{e} &$F_{\rm unabs}$\tablenotemark{f} &($C$ or $\chi^2_{\nu}$)/dof\tablenotemark{g} \\ \hline
{\it Chandra} (1$''$) &4.7$^{+3.6}_{-2.8}$&1.72$^{+0.55}_{-0.63}$&9.8$^{+9.1}_{-5.8}$&0.14$^{+0.06}_{-0.04}$&13$^{+454}_{-12}$&6.7$^{+0.9}_{-0.8}$&32$^{+190}_{-21}$&28.2/29 \\
{\it XMM-Newton} (20$''$) &2.2$^{+1.0}_{-0.7}$&1.35$^{+0.16}_{-0.17}$&$22\pm5$&0.18$^{+0.04}_{-0.03}$&2$^{+6}_{-1}$&17.8$\pm$0.8&26$^{+8}_{-4}$&1.27/80 \\
{\it Chandra} (20$''$) &2.4$^{+3.1}_{-1.9}$&1.32$^{+0.50}_{-0.69}$&18$^{+18}_{-12}$&0.22$^{+0.13}_{-0.09}$&0.5$^{+22}_{-0.4}$&$16\pm2$&22$^{+31}_{-5}$&1.03/82 \\
{\it Combined} (20$''$) &1.9$^{+0.8}_{-0.6}$&1.33$^{+0.16}_{-0.17}$&$21\pm5$&0.19$^{+0.04}_{-0.03}$&0.9$^{+2.8}_{-0.6}$&$17.3\pm0.7$&24$^{+5}_{-3}$&1.18/167 \\
\hline\hline
\end{tabular}
\tablecomments{The errors shown represent 90\% confidence intervals.}\\ 
\tablenotetext{\rm a}{~Radius of extraction region is given in parentheses.}
\tablenotetext{\rm b}{~Spectral flux in units of 10$^{-6}$ photons cm$^{-2}$ s$^{-1}$ keV$^{-1}$ at 1 keV.}
\tablenotetext{\rm c}{~BB temperature in keV.}
\tablenotetext{\rm d}{~Projected emitting area for BB model in units of km$^{2}$ at the distance of 2.5 kpc.}
\tablenotetext{\rm e}{~Absorbed flux in the 0.3--8 keV, in units of $10^{-14} \, \rm erg\, cm^{-2} \, s^{-1}$.}
\tablenotetext{\rm f}{~Unabsorbed flux in the 0.3--8 keV, in units of $10^{-14} \, \rm erg\, cm^{-2} \, s^{-1}$.}
\tablenotetext{\rm g}{~Best-fit $C$-statistic value for {\it Chandra} observation (1$''$ radius aperture) or reduced $\chi^2$ value (other cases).}
\end{table*}

\begin{figure}[h!]
\centering
\includegraphics[scale=0.32,angle=-90]{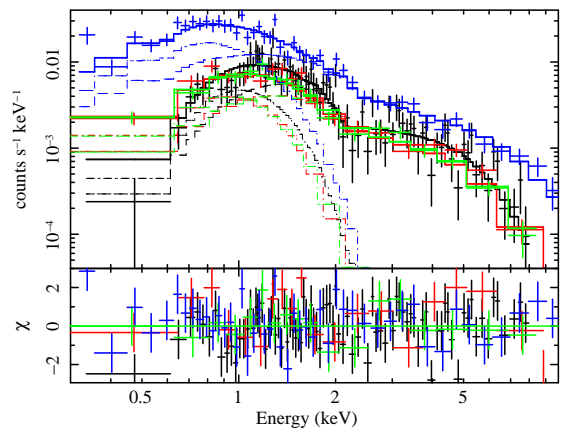}
\caption{ACIS-I ({\it black}), EPIC MOS1 ({\it red}), MOS2 ({\it green}), and PN ({\it blue}) spectra of the pulsar region (20$''$ radius) fitted with the absorbed PL (dashed line) + BB (dash-dotted line) model.}
\label{fig9}
\end{figure}

The photon index and temperature of the PL+BB model inferred from the {\it XMM-Newton} and {\it Chandra} observations are consistent within their uncertainties. On the other hand, the absorbed flux estimated from the {\it XMM-Newton} data is a factor of three larger than that estimated from the {\it Chandra} data due to the larger extraction region (20$''$ radius), chosen  for {\it XMM-Newton} data, which contains a significant PWN contribution. Therefore, we extracted the spectra using the same source region (20$''$ radius) for the {\it Chandra} observation to compare with the spectral results obtained from the {\it XMM-Newton} observation. The region includes 873 source counts, which is 83\% of the total counts. Now both fits are consistent with each other (see Table.~\ref{tab2}). Finally, the {\it Chandra} (20$''$ radius aperture) and {\it XMM-Newton} data were fitted jointly with the PL+BB model, which still  fits satisfactorily while the uncertainties of the best-fit parameters are significantly reduced compared to the individual fits  (see Table~\ref{tab2} and Figure~\ref{fig9}). 

%Fig.9

\subsubsection{PWN Spectrum}
\begin{figure}[h!]
\centering
\includegraphics[scale=0.32,angle=90]{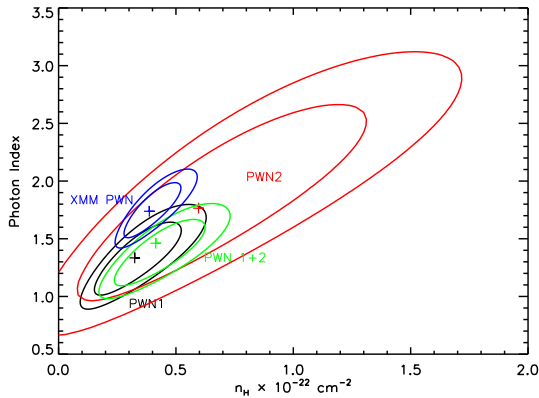}
\caption{Confidence contours (68\% and 90\%) in the $n_{\rm H}-\Gamma$ plane for the PWN spectrum measured with {\it XMM-Newton} ({\it blue}) and the PWN1 ({\it black}), PWN2 ({\it red}), and PWN\,1+2({\it green}) spectra measured with {\sl Chandra}. }
\label{fig10}
\end{figure}

We extracted the ACIS spectra from the PWN1 and PWN2 regions (see Fig.~\ref{fig1}). For the PWN1's spectrum, we excluded the pulsar region (1$\farcs$5 radius aperture) and used the absorbed PL model in the energy range of 0.3--8 keV. The spectrum was binned with minimum of 10 counts per bin. The fit ($\chi^2_{\nu}=1.3$ for 24 dof) gives $n_{\rm H, 21}$ = 3.2$^{+2.3}_{-1.4}$, $\Gamma$ = 1.31$^{+0.37}_{-0.33}$, and $F^{\rm abs}_{\rm pwn1}=(6.0\pm0.9) \times 10^{-14}$ erg cm$^{-2}$ s$^{-1}$. The spectrum of PWN2 was also binned with minimum of 10 counts per bin and fitted with the absorbed PL model in the same energy range. We used the $C$-statistic because of small number of counts in the region. The PL fit to the spectrum of the PWN2 gives somewhat larger $\Gamma$ = 1.8$^{+1.0}_{-0.9}$; however, the uncertainties are large as one can see from the confidence contours shown in Figure~\ref{fig10}. We also extracted and fitted the spectrum from both regions combined, since the difference in slopes of both spectra is not statistically significant. The fit gives $n_{\rm H, 21}$ = 3.7$^{+2.0}_{-1.3}$, $\Gamma=1.37\pm0.22$, and the absorbed flux is $F^{\rm abs}_{\rm pwn1+2}=(7\pm1)\times10^{-14}$ erg cm$^{-2}$s$^{-1}$ in the 0.3--8 keV band, while the unabsorbed flux is $F^{\rm unabs}_{\rm pwn1+2}=9^{+2}_{-1}\times10^{-14}$ erg cm$^{-2}$s$^{-1}$ in the same energy band. It corresponds to the luminosity $L_{\rm PWN 1+2} \sim 7\times10^{31} d^2_{2.5}$ erg s$^{-1}$, i.e., about 130\% of the pulsar's non-thermal luminosity estimated from the same observation.

%Fig.10

For the PWN region on the I1 chip (see Fig.~\ref{fig1}), we binned the spectrum with minimum 25 counts per bin and fitted with the absorbed PL model in the 0.3--8 keV energy range. We fixed the hydrogen column density to the value $n_{\rm H,21}=4.7$, found from the PL+BB model for pulsar obtained from the {\it Chandra} data (see Table 2). The fit ($\chi^2_{\nu}=0.66$ for 10 dof) gives $\Gamma = 2.5^{+1.8}_{-0.9}$ and $\mathcal{N}_{\Gamma}=8^{+6}_{-4}\times 10^{-6}$ photons cm$^{-2}$ s$^{-1}$ keV$^{-1}$ at 1 keV. The absorbed flux in the same energy range is $1.3^{+0.9}_{-0.7} \times 10^{-14}$ erg cm$^{-2}$ s$^{-1}$, while the unabsorbed flux is $4^{+10}_{-1} \times 10^{-14}$ erg cm$^{-2}$ s$^{-1}$. The spectrum of the PWN region on the I2 chip was binned with minimum of 50 counts per bin and fitted with a single absorbed PL model in the energy of 0.3--8 keV. We fixed the hydrogen column density of $n_{\rm H,21}=4.7$. The fit was acceptable ($\chi^2_{\nu}=0.7$ for 18 dof) with $\Gamma = 1.7 \pm 0.4$ and $\mathcal{N}_{\Gamma}=(1.5\pm0.5)\times 10^{-5}$ photons cm$^{-2}$ s$^{-1}$ keV$^{-1}$ at 1 keV. The absorbed flux in the same energy range is $(7\pm3) \times 10^{-14}$ erg cm$^{-2}$ s$^{-1}$, while the unabsorbed flux is $(9\pm2) \times 10^{-14}$ erg cm$^{-2}$ s$^{-1}$. 

%Table.3
\begin{table*}
\footnotesize
\center
\caption{Spectral properties of the PWN from the {\it Chandra} and {\it XMM-Newton} observations\label{tab3}}
\vspace{0.5cm}
\setlength{\tabcolsep}{1mm}
\begin{tabular}{ccccccc}\hline\hline
Region&$n_{\rm H,21}$&$\Gamma$&$\mathcal{N}$\tablenotemark{a}&$F_{\rm abs}$\tablenotemark{b}&$F_{\rm unabs}$\tablenotemark{c}&($C$ or $\chi^2_{\nu}$)/dof\tablenotemark{d} \\\hline
{\it Chandra}&&&&&&\\
PWN1&3.2$^{+2.3}_{-1.4}$&1.31$^{+0.37}_{-0.33}$&8.5$^{+4.7}_{-2.9}$&$6.0\pm0.9$&$7\pm1$&1.3/24 \\
PWN2&6.0$^{+7.9}_{-5.5}$&1.8$^{+1.0}_{-0.9}$&3.7$^{+9.5}_{-3.7}$&1.3$^{+0.5}_{-0.4}$&2.2$^{+4.2}_{-0.8}$&1.04/5 \\
PWN\,1+2&3.7$^{+2.0}_{-1.3}$&$1.37\pm0.22$&12$^{+6 }_{-4}$&$7\pm1$&9$^{+2}_{-1}$& 0.90/31 \\
PWN\,I1&4.7(fixed)&2.5$^{+1.8}_{-0.9}$&8$^{+6 }_{-4}$&1.3$^{+0.9}_{-0.7}$&4$^{+10}_{-1}$&0.66/10\\
PWN\,I2&4.7(fixed)&$1.7\pm0.4$&$15\pm5$&$7\pm3$&$9\pm2$&0.70/18\\
{\it XMM-Newton}&&&&&&\\
PWN&3.9$^{+1.5}_{-1.1}$&1.74$^{+0.27}_{-0.24}$&123$^{+46}_{-31}$&$50\pm6$&75$^{+14}_{-9}$&0.94/61 \\
\hline\hline
\end{tabular}
\tablecomments{The errors shown represent 90\% confidence intervals.}\\
\tablenotetext{\rm a}{~Spectral flux in units of 10$^{-6}$ photons cm$^{-2}$ s$^{-1}$ keV$^{-1}$ at 1 keV.}
\tablenotetext{\rm b}{~Absorbed flux in the 0.3--8 keV, in units of $10^{-14} \, \rm erg\, cm^{-2} \, s^{-1}$.}
\tablenotetext{\rm c}{~Unabsorbed flux in the 0.3--8 keV, in units of $10^{-14} \, \rm erg\, cm^{-2} \, s^{-1}$.}
\tablenotetext{\rm d}{~$C$-statistic value for PWN2, $\chi^2$ values for other regions.}
\end{table*}

\begin{figure}[h!]
\centering
\includegraphics[scale=0.32,angle=-90]{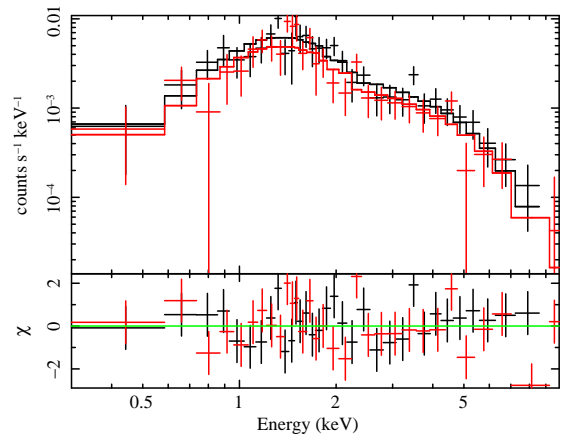}
\caption{EPIC MOS1 ({\it black}) and MOS2 ({\it red}) spectra of the PWN shown in Figure~\ref{fig3}.}
\label{fig11}
\end{figure}

Figure~\ref{fig11} shows the MOS1 and MOS2 spectra of the region shown in Figure~\ref{fig3}, binned with minimum 50 counts per bin. The region is much larger than that of PWN1+PWN2 of {\sl Chandra} (see Figures \ref{fig1} and \ref{fig3}). The absorbed PL fit is good ($\chi^2_{\nu}$ = 0.94 for 61 dof). The best-fit hydrogen column density  and  photon index are $n_{\rm H, 21}$ = 3.9$^{+1.5}_{-1.1}$ and $\Gamma$ = 1.7$^{+0.3}_{-0.2}$, respectively. The absorbed flux is $F^{\rm abs}_{\rm pwn} = (5.0\pm0.6) \times 10^{-13}$ erg cm$^{-2}$s$^{-1}$ in 0.3--8 keV. The unabsorbed flux $F^{\rm unabs}_{\rm pwn} = 7.5^{+1.4}_{-0.9} \times 10^{-13}$ erg cm$^{-2}$s$^{-1}$. The results of the fits, summarized in Table~\ref{tab3}, 
suggest that the PWN spectrum softens with the distance from the pulsar. The combined compact+extended PWN flux is $F^{\rm unabs}_{\rm pwn}\approx 9 \times 10^{-13}$ erg cm$^{-2}$s$^{-1}$ which corresponds to the the luminosity $L_{\rm pwn}=6.7\times10^{32} d^2_{2.5}$ erg s$^{-1}$ $\approx 2.8L_{\rm psr}$. 

 %Fig.11

\subsubsection{Timing Analysis}
We searched for pulsations in the PN data (time resolution of 6 ms) extracted from the $r=20''$ circle centered at the pulsar position, which contains 939 counts, including 15\% from the background, in the 0.3 -- 1.1 keV band, where the BB component is dominant (see Figure~\ref{fig9}). From the same region in the 1.1 -- 10 keV band, we extracted 1473 counts, including 21\% from the background. The photon arrival times were corrected to the solar system barycenter using the SAS {\it barycen} tool. The PN observation started at the epoch 55,037.63605 MJD and continued for the time $T_{\rm span}= 55.5$ ks. We used the radio timing ephemeris reported by \citet{Camilo2004} ($f=6.020\,167\,772\,6(4)$ s$^{-1}$ and $\dot{f}=-1.305\,395(4)\times10^{-11}$ s$^{-2}$ for the epoch of 52,921.0 MJD) to calculate the predicted frequency at our epoch. There was a glitch at the epoch of 52,021$\pm$16 MJD (1$\sigma$ uncertainties), but it was before the ephemeris was measured. We assumed that there were no glitches after that. The predicted frequency for our observation is $f_{\rm predict}= 6.017\,780\,858(7)$ Hz. 

We applied the $Z^2_1$ test \citep{Buccheri1983,Pavlov1999} to search for the pulsed signal in the band $6.017\,780\pm0.000\,010$ Hz near the predicted frequency, with a step of 0.1 $\mu$Hz $\approx 0.006 T_{\rm span}^{-1}$. 
We found the maximum $Z^2_{\rm 1, max} = 40.5$ at $f = 6.017\,780\,1~\rm{Hz} \pm 1.6~\mu \rm{Hz}$, i.e., the pulsations are detected with a $6\sigma$ significance in the 0.3--1.1 keV energy range\footnote{The $1\sigma$ frequency uncertainty is calculated as $\delta f=
3^{1/2}\pi^{-1} T_{\rm span}^{-1} (Z_{1,\rm max}^2)^{-1/2}$. This equation can be derived using the method outlined in \S\,2.4 of \citet{Bretthorst1988}. The numerical factor, $\sqrt{3}/\pi\approx 0.55$, was confirmed in Monte Carlo simulations by A.\ Kaurov (2011, priv.\ comm.).}.
The maximum value of $Z^2_1$ in the 1.1--10 keV energy range is only 7.9 at $f=6.017\,781\,9~\rm{Hz} \pm 3.5~\mu \rm{Hz}$, which corresponds to $2.3\sigma$ significance. 

\begin{figure}[h!]
\centering
\includegraphics[scale=0.4,angle=0]{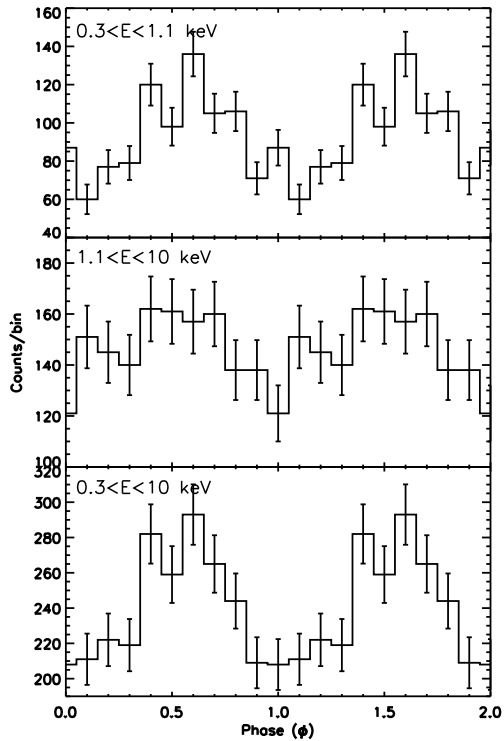}
\caption{Pulse profiles of J1357 extracted from the EPIC PN data in three energy bands.}
\label{fig12}
\end{figure}

Using the two epochs (55,037.63605 MJD from our observation and 52,921.0 MJD from the radio) and two frequency values, we can estimate the frequency derivative, $\dot{f}=-1.3056(7) \times 10^{-11}$ s$^{-2}$, which is consistent with $\dot{f}$ determined from the radio. Figure~\ref{fig12} shows the pulse profiles with 10 phase bins, with the estimated pulsed fraction (the ratio of the number of counts above the minimum level to the total number of counts) $p = 36\% \pm 5\%$ in 0.3--1.1 keV, $18\% \pm 4\%$ in 1.1--10 keV, and $14\%\pm 3\%$ in 0.3--10 keV.
We estimated the error of pulsed fraction as $\delta p=\sqrt{2/N}$ \citep{Linsley1975}. After correcting for the background, we obtain the intrinsic pulsed fraction $p_{\rm int} = 42\% \pm 5\%$, $23\% \pm 4\%$, and $17\%\pm 3\%$, respectively. 
Thus, we conclude that the X-ray emission from J1357 is undoubtedly pulsed, although the pulsed fraction is not as high as reported by \citet{Zavlin2007}. Interestingly, the pulsed fraction is higher at lower photon energies, where the thermal component dominates. Such strong pulsations of the thermal emission are likely due to the anisotropy of atmospheric radiation in the strong magnetic field of this pulsar.

%Fig.12

\section{Discussion}

Our {\sl Chandra} ACIS and {\sl XMM-Newton } EPIC observations have shown that the PWN of J1357 is much more extended and luminous  than it could be seen from the earlier observations. We have also found that the thermal component of pulsar emission is strongly pulsed, while the pulsation of the nonthermal component, which dominates at higher energies, are much weaker. Below we discuss the physical implications of our findings, compare the J1357 PSR/PWN properties to those of other similar pulsars observed in X-rays, and  investigate the relation between J1357, HESS~J1356, and putative host SNR G309.8--2.6.

\subsection{Pulsar}

 The pulsar spectrum cannot be satisfactory fitted by simple one-component models, such as absorbed PL or BB (see \S2.2.1).  The two-component BB+PL model fits the spectrum much better, although some residuals are still present in the combined fits. The best-fit photon index for the PL component, $\Gamma\simeq1.7\pm0.6$, is marginally larger than $\Gamma\simeq1.3\pm0.3$ of the compact PWN (region PWN1). The non-thermal luminosity is $L_{\rm nonth}\approx5\times10^{31}$ erg s$^{-1}$ in the energy range 0.3--8 keV. The PL component may partly come from an unresolved PWN. This could explain the weak pulsations above 1 keV (see \S2.2.3), in contrast to, for instance, the Vela pulsar (which is well resolved from the surrounding PWN), whose pulsed fraction grows with energy reaching 62\% above 1.8 keV \citep{Sanwal2002}. 
Our timing analysis limits the luminosity of the unresolved PWN to $\lesssim 77\%$ of the point source luminosity in 1.1-10 keV.

 Unlike the hard non-thermal component, the soft thermal component should not be strongly contaminated by the PWN emission since it clearly dominates at $\lesssim1$ keV. Therefore, the best-fit BB temperature, $T\approx1.9-2.7$ MK, radius, $R\sim 0.3-1.1$ km, and bolometric luminosity, $L_{\rm bol}\sim
%\approx 8.6
2\times10^{31}$ erg s$^{-1}$, are reasonably accurately determined (see Table~\ref{tab2}).
The values are similar to those of other Vela-like pulsars. Although the temperature is a factor of two higher than expected from the standard NS cooling, and the emitting area radius is a factor of ten smaller then the classical NS radius, the discrepancies can be attributed to the use of the simplistic BB model. 
Indeed, the fit with the hydrogen atmosphere models \citep[e.g.,][]{Pavlov1995}, described in Section 2.2.1, gives a lower temperature,
$T_{\rm eff}^\infty = 0.70$--0.78 MK at $R^\infty = 13$ km,
at the same bolometric luminosity.
The measured bolometric luminosity is a factor of $\sim 
3$ lower than predicted by the theoretical cooling curve for low-mass NSs ($M_{\rm NS} = 1.3 M_\odot$) without superfluidity; it is consistent with the cooling curves for heavier NSs (e.g., $M_{\rm NS}=1.5$--$1.6\, M_\odot$) for some superfluidity models \citep{Yakovlev2004}.
 
Very surprising is the high ($\approx40$\%) pulsed fraction associated with the thermal component. In general, one should not expect such strong pulsations from a non-uniformly heated NS surface with dipolar magnetic field for a blackbody-like angular distribution.
However, there have been recent reports of even stronger thermal pulsations in strongly magnetized young pulsars, such as PSR J1119--6127 ($74\%\pm14\%$; Gonzalez et al.\  2005) and PSR J1718--3718 ($52\%\pm13\%$; Zhu et al. 2011). This suggests that the strong pulsations are caused by the combined effect of the anisotropic temperature distribution and strong anisotropy of emitted radiation in high magnetic fields \citep{Pavlov1994}. An alternative possibility is that part of emission interpreted as thermal from our BB+PL
and NSA+PL fits is in fact an additional, softer non-thermal component, which is masked at higher energies by the harder, unresolved PWN emission.  This unaccounted component might be the reason for the remaining residuals of the PL+BB and NSA+PL fits. However, higher quality data are needed to test this hypothesis.

\subsection{PWN}

At the plausible distance of 2.5 kpc, the unabsorbed X-ray luminosity of the compact PWN, $L_{\rm pwn1}\approx 5 \times 10^{31}$ ergs s$^{-1}$ in the 0.3-8 keV band, corresponds to the X-ray efficiency, $\eta_{\rm pwn1}\equiv L_{\rm pwn1}/\dot{E} \sim 1.7\times10^{-5}$, lower than those of most Vela-like pulsars (see Kargaltsev et al.\ 2007; KP08). This suggests  that either the distance is underestimated or, in addition to $\dot{E}$ and $\tau$, $\eta_{\rm pwn1}$ strongly depends on other factors (e.g., the pulsar's speed and the angle between the spin and magnetic axes). Note, however, that the efficiency of the entire detected PWN is higher by a factor of 15, $\eta_{\rm pwn}\sim 2.6\times10^{-4}$.

The spectral slope of the compact PWN, $\Gamma_{\rm pwn1} =1.3\pm0.3$, is similar to those of PWNe around Vela-like pulsars \citep[listed in Table 2 of][]{Karga2007c}. The PWN spectrum apparently softens with the distance from the pulsar, becoming  $\Gamma\approx1.7-1.8$ for the narrow feature (PWN2) and the large-scale PWN, which suggests synchrotron cooling of the outflow. The degree of spectral softening is similar to that seen for the extended tail of  PSR J1509--5850, where the photon index changes from $\Gamma=1.8\pm0.3$ for PWN in the vicinity of the pulsar to $2.4\pm0.4$ for the extended tail \citep{Karga2008a}.

 Our X-ray images reveal the morphology of the PWN on different  angular scales. The observed  morphology does not easily  fit in either bow-shock or torus-jet category  (see KP08).  The bright compact PWN1 extends up to $\simeq 10''$ from the pulsar and  has rather amorphous, somewhat asymmetric  morphology.  The diffuse emission is noticeably brighter northeast of the pulsar compared to the opposite side, although some emission is still clearly discernible within the PWN1 region southwest of the pulsar. We find no evidence of  structures that could be interpreted as a torus associated with the termination shock (TS) in the pulsar wind, which may suggest that the TS occurs at an angular distance of several arcseconds or less, corresponding to $r_s\lesssim4\times10^{16}d_{2.5}$ cm (versus $1\times10^{17}$ cm in the Vela PWN; Helfand et al.\ 2001). If the pulsar is moving subsonically inside the hot SNR medium, we can use the above estimate to obtain a lower limit on the ambient pressure at $p_{\rm amb}\sim \dot{E}  (4\pi c r_s^2)^{-1}\gtrsim5\times 10^{-9}(r_s/4\times10^{16}~{\rm cm})^{-2}$ dyn cm$^{-2}$, which is close to the maximum pressure one can expect in a 10 kyr old SNR \citep{Karga2009, Bamba2010}.
The high pressure required, and the asymmetry of the compact PWN, elongated approximately along the northeast-southwest direction, might be indicative of the TS being also asymmetric, i.e., being farther away from the pulsar on the northeast side than on the southwest side. Such situation may occur if the pulsar is moving fast through the surrounding medium in the southwest direction, and the ram pressure of the medium compresses the PWN in front of the pulsar. If the pulsar's speed exceeds the sound speed in the medium, $c_s=(5 kT/3 \mu m_{\rm H})^{1/2}=12\, \mu^{-1/2} T_4^{1/2}$ km s$^{-1}$, a bowshock is formed. The appearance of a bowshock PWN depends on the Mach number, uniformity of the medium, and intrinsic anisotropy of the pulsar wind.  X-ray PWNe around supersonically moving pulsars often exhibit a bright ``bullet'' in the vicinity of the pulsar and a much more extended faint tail in the direction opposite to that of the pulsar motion; however, more complex structures have been seen in some cases (e.g., the Guitar and Eel PWNe associated with  B2224+65 and J1826--1256, respectively; \citet{Johnson2010, Roberts2007}).
  
 The adjacent PWN2 region shown in Figure 1 encompasses the fainter narrow feature attached to the compact PWN. Interestingly, this narrow feature bends sharply at about $23''$ from the pulsar. Although there is no doubt that the feature is part of the PWN, the origin of the feature is not clear. For instance, it could be a pulsar jet. Indeed, some of pulsar jets are known to show extreme bending (e.g., the outer jet of the Vela pulsar; \citet{Pavlov2003}).
The apparent lack of a similarly looking counter-jet should not be surprising since the two jets can differ significantly both in shape and surface  brightness due to the Doppler boost and/or proper motion effects  (e.g., \citet{Pavlov2003}).
On the other hand, the ratio of the compact PWN luminosity to that of the putative jet, $L_{\rm pwn1}/L_{\rm pwn2}\sim5$, is noticeably larger than that in the Vela and Crab PWNe but comparable to that of the PWN around PSR B1706--44 \citep{Romani2005}. If this interpretation is taken at the face value, one could expect the pulsar to be moving northeast (in general, jets tend to be co-aligned with the pulsar's direction of motion--cf.\ the Vela pulsar; \citet{Pavlov2003}).

Alternatively, the narrow feature could be a part (e.g., an inner channel) of the pulsar tail (see, e.g., \citet{Karga2008}
and references therein). In this case the pulsar would be moving in the opposite (southwest) direction. This interpretation is  in line with the above-described asymmetry of the compact PWN whose brighter part could, in this case, indeed be the ``bullet'' associated with the termination shock deformed by the ram pressure of the oncoming ISM.  However, the sharp bending and the lack of a fainter, much more extended tail (cf.\ J1509--5850 tail; \citet{Karga2008}),
and the emission west-southwest of the compact PWN, are at odds with this interpretation\footnote{The diffuse emission seen in the PMN 4.85 GHz image north of the pulsar (panel e in Fig.~6) could, in principle, be the radio counterpart of the extended pulsar tail. Deeper radio observations are required to test this hypothesis.}. 

Finally, the narrow feature could be a part of the bow associated with the forward shock (see e.g., the Geminga PWN; Pavlov et al.\ 2006, 2010) with the pulsar moving southeast.  In this case the observed large-scale emission can also be associated with the asymmetric bowshock similar to that seen around  PSR J1826--1256 \citep{Roberts2007}. Somewhat puzzling may be the lack of any X-ray emission within the bow interior; however, this is also the case in the Guitar and Eel bowshocks. To conclude, with the data in hand we cannot univocally interpret the morphology of the X-ray PWN. Measuring the pulsar's proper motion should provide decisive  information allowing one to discriminate between the above interpretations. Assuming a typical pulsar speed of 400 km s$^{-1}$ \citep{Hobbs2005}, this should be possible to accomplish in several years with VLBI.

%Table.4
\begin{table*}
\footnotesize
\center
\caption{Properties of candidate TeV plerions and their parent pulsars.\label{tab4}}
\vspace{0.5cm}
\setlength{\tabcolsep}{1mm}
\begin{tabular}{lccccccccccc}
\hline\hline
HESS ID &  $f_{\gamma}$\tablenotemark{a} & TeV size &  Offset\tablenotemark{b}  & PSR & $\dot{E}_{36}$\tablenotemark{c} & $\tau$ & Dist. & $L_{\gamma}/\dot{E}$ &  $L_{X}/L_{\gamma}$ &  $\mathcal{B}_{\rm acis}$\tablenotemark{d}   \\
     &   & pc &  &  &  & kyr & kpc & & &\\\hline
  J1809$-$193 & 0.3 & $20$ & $8'$ &  J1809--1917 & 1.8  & 50 & 3.5 & 1\% & $2\%$ & 8  \\
 Vela X  & 0.75 & $5$ & $30'$ & J0835--4510 & 6.9 & 11 & 0.3 & 0.002\% & $90\%$ & -- \\
 J1825$-$137 & 0.48 & $70$ & $10'$ & J1826--1334 & 2.8 & 21 & 3.5 & $4\%$ & $0.3\%$ & 7.5  \\
 J1356$-$645 & 0.79 & $40$ & $8'$  & J1357--6429 & 3.1 & 7.3 & 2.5 & $0.2\%$ & $10\%$ & 19 \\
 \hline\hline
\end{tabular}
\tablenotetext{\rm a}{~Unabsorbed $\gamma$-ray flux (1--10 TeV) in units of $10^{-11}$ ergs s$^{-1}$ cm$^{-2}$.}
\tablenotetext{\rm b}{~Offset of the TeV source center from the pulsar.} 
\tablenotetext{\rm c}{~Pulsar spin-down power, $\dot{E}$, in units of  $10^{36}$ ergs s$^{-1}$.}
\tablenotetext{\rm d}{~Average surface brightness  measured from {\sl Chandra} ACIS images, in counts ks$^{-1}$ arcmin$^{-2}$.}
\end{table*}

\subsection{The Nature of the VHE source}

 Given the lack of other promising counterparts and the location of J1357 within the HESS~J1356 extent, it seems very plausible that the two should be related in some way.  Yet the nature of HESS~J1356 has not been firmly  established.  The TeV emission could be attributed to either the host SNR or to the relic PWN of J1357.  To date, several well-known SNRs have been detected in the TeV band \citep[e.g.,][]{Aha2004}. In the cases of resolved SNRs, the TeV emission was associated with the SNR shell, and the sizes of the SNRs in TeV images were similar to those of the non-thermal radio shells indicating that both emission components are powered by particles accelerated in the forward shock.  Although no shell is seen in the radio images of HESS~J1356, the radio emission is mainly seen  near the center of HESS~J1356 (see Fig.~\ref{fig6}). Thus, in this case the TeV emission does not appear to be associated with the nonthermal shell, and the shell itself is not seen in the radio. 

An alternative, more plausible, interpretation of the TeV and radio emission could be a relic PWN. Many extended TeV sources  neighbor young Vela-like pulsars, sometimes offset up to $10'$--$20'$ from the center of TeV emission (see KP10 for recent review). Furthermore, recent X-ray observations of the Vela pulsar region  (Vela X; Mori et al.\ 2008),  PSR~J1826--1334 (Gaensler et al.\ 2003; Pavlov et al.\ 2008) and PSR~J1809--1917 \citep{Karga2007} have provided convincing evidence that the TeV sources are connected to the parent pulsars through faint asymmetric X-ray nebulae\footnote{There remains a large fraction of relic PWN candidates where only bright compact PWN is seen X-rays while the extended component is not seen perhaps due to is faintness and/or rapid cooling of the outflow.} (see Table~\ref{tab4}). Indeed, the {\sl Chandra} observations have demonstrated that in the above three examples the PWNe consist of compact (0.1--0.8 pc) bright cores and  fainter asymmetric, more extended (2--8 pc) components. The  offsets and the asymmetries of the X-ray PWNe could be created by the reverse SNR shock that had propagated through the nonhomogeneous SNR interior and reached one side of the PWN sooner than the other side \citep{Blondin2001}. This scenario could also account for the similarly asymmetric, offset TeV emission (e.g., de Jager \& Djannati-Ata\"{\i} 2009).  However, the physical origin of the TeV emission still remains under debate. It can be produced via the Inverse Compton Scattering (ICS) of the relativistic pulsar-wind electrons on photons from the omnipresent cosmic microwave background (CMB), galactic IR background, and IR photons from local star-forming regions and warm dust clouds. Alternatively, the TeV photons can be produced as a result of $\pi^{0}\rightarrow\gamma + \gamma$ decay, with $\pi^{0}$ being produced when the relativistic protons from the pulsar wind interact with the ambient matter (e.g., Horns et al.\ 2007). In the case of HESS~J1356, the offset is well within the range observed in other relic PWN candidates (see KP10). 
The ratio of the TeV luminosity to the pulsar's
spin-down power, $L_{\rm TeV}/\dot{E} \sim 0.002$, and the ratio of X-ray to $\gamma$-ray 
(1--10 TeV) 
PWN luminosities, $L_X/L_{\rm TeV}\sim 0.1$, are  
similar to those of other relic plerions.

  Alternatively, the relic PWN could be left behind the fast moving pulsar if the latter is moving in the northeast direction (see above). At a typical speed of 400 km s$^{-1}$, the pulsar would move by $5\farcm 6$ in 10 kyrs, which is 
consistent with the offset between the pulsar and the center of HESS~J1356. So far, there have been no reports of VHE emission from pulsar tails. It would be interesting to establish the presence of the TeV emission from an extended  pulsar tail because this would help to break the degeneracy in interpreting the nature of the VHE emission from the crushed PWNe, for which both leptonic (Inverse Compton) and hadronic ($\pi^0$ decay) TeV emission mechanisms are currently being debated. If the $\pi^0$ decay is the dominant process in the crushed PWNe, we would expect pulsar tails to be significantly fainter in the TeV  because high-speed pulsars move in low density media outside their host SNRs. On the other hand,  in the leptonic scenario, crushed plerions and pulsar tails should have comparable TeV luminosities. Regardless of the nature of the relic PWN and the mechanism responsible for the VHE emission, one can expect that the observed extended radio emission within HESS~J1356 may also be produced by the pulsar wind if the electron SED extends to sufficiently low energies.  Better quality radio images, more complete multiwavelength spectrum, and robust multizone pulsar wind models are required to assess the origin of the radio emission.

\acknowledgements
 
  The work on this project was partly supported by Chandra award GO0-11070X, NASA grants NNX09AT11G, NNX09AC81G and NNX09AC84G, and National Science Foundation grants AST09-08733 and AST09-08611. The work by G.G.P. and Y.A.S. was partly supported by the Ministry of Education and Science of the Russian Federation (contract 11.G34.31.0001). We thank Alexander Kaurov for his help with simulations of timing uncertainties.


\begin{thebibliography}{109}
\bibitem[Abramowski et al.(2011)]{Abramowski2011} Abramowski, A., et al. 2011, preprint (arXiv1108.2855)
\bibitem[Aharonian et al.(2004)]{Aha2004} Aharonian,~F.~A., et al. 2004, Nature, 432, 75
\bibitem[Bamba et al.(2010)]{Bamba2010} Bamba,~A., Mori,~K. \& Shibata,~S. 2010, ApJ, 709, 507
\bibitem[Blondin et al.(2001)]{Blondin2001} Blondin,~J.~M., Chevalier,~R.~A., \& Frierson,~D.~M. 2001, ApJ, 563, 806 
\bibitem[Bretthorst (1988)]{Bretthorst1988} Bretthorst, G.\ L. 1988, Bayesian Spectrum Analysis and Parameter Estimation (Springer: New York)
\bibitem[Buccheri et al.(1983)]{Buccheri1983} Buccheri,~R., et al. 1983, A\&A, 128, 245
\bibitem[Bucciantini et al.(2005)]{Bucciantini2005} Bucciantini,~N., Amato,~E. \& Del Zanna,~L. 2005, A\&A, 434, 189
\bibitem[Camilo et al.(2004)]{Camilo2004} Camilo,~F., et al. 2004, ApJ, 611, L25
\bibitem[Cash(1979)]{Cash1979} Cash,~W. 1979, ApJ, 228, 939
\bibitem[Cordes \& Lazio(2002)]{Cordes2002} Cordes,~J.~M., \& Lazio,~T.~J.~W. 2002, preprint (astro-ph/0207156)
\bibitem[de Jager \& Djannati-Ata\"{\i}(2008)]{deJager2008} de Jager,~O.~C. \& Djannati-Ata\"{\i},~A. 2009, in Neutron Stars and Pulsars: 40 Years After Their Dscovery, ed. W. Becker (Astrophys. Space Sci. Lib.357; Berlin: Springer), 451
\bibitem[de Luca et al.(2005)]{deLuca2005} de Luca,~A., Caraveo,~P.~A., Mereghetti,~S., Negroni,~M., \& Bignami,~G.~F. 2005, ApJ, 623, 1051
\bibitem[Duncan et al.(1997)]{Duncan1997} Duncan,~A.~R., Steward,~R.~T., Haynes,~R.~F., \& Jones,~K.~L. 1997, MNRAS, 287, 722
\bibitem[Esposito et al.(2007)]{Esposito2007} Esposito,~P., Tiengo,~A., de Luca,~A., \& Mattana,~F. 2007, A\&A, 467, L45
\bibitem[Gaensler \& Slane(2006)]{Gaensler2006} Gaensler,~B.~M., \& Slane,~P.~O. 2006, ARA\&A, 44, 17
\bibitem[Gaensler et al.(2003)]{Gaensler2003} Gaensler,~B.~M., Schulz,~N.~S., Kaspi,~V.~M., Pivovaroff,~M.~J., \& Becker,~W.~E. 2003, ApJ, 588, 441
\bibitem[Gonzalez et al.(2005)]{Gonzalez2005} Gonzalez,~M.~E., Kaspi,~V.~M., Camilo,~F., Gaensler,~B.,~M., \& Pivovaroff,~M.,~J. 2005, ApJ, 630, 489 
\bibitem[Griffith \& Wright(1993)]{Griffith1993} Griffith,~M.~R., \& Wright,~A.~E. 1993, AJ, 105, 1666
\bibitem[Helfand et al.(2001)]{Helfand2001} Helfand,~D.~J., Gotthelf,~E.~V., \& Halpern,~J.~P. 2001, ApJ, 556, 380  
\bibitem[Hobbs et al.(2005)]{Hobbs2005} Hobbs,~G., Lorimer,~D.~R., Lyne,~A.~G., \& Kramer,~M. 2005, MNRAS, 360, 974
\bibitem[Horns et al.(2007)]{Horns2007} Horns,~D., Aharonian,~F., Hoffmann,~A.~I.~D., \& Santangelo,~A. 2007, Ap\&SS, 309, 189  
\bibitem[Johnson \& Wang(2010)]{Johnson2010} Johnson,~S.~P., \& Wang,~Q.~D. 2010, MNRAS, 408, 1216
\bibitem[Kargaltsev \& Pavlov(2007)]{Karga2007} Kargaltsev,~O., \& Pavlov,~G.~G. 2007, ApJ, 670, 655 
\bibitem[Kargaltsev \& Pavlov(2008)]{Karga2008} Kargaltsev,~O., \& Pavlov,~G.~G. 2008, AIP proc., 983, 171 
\bibitem[Kargaltsev \& Pavlov(2010)]{Karga2010} Kargaltsev,~O., \& Pavlov,~G.~G. 2010, AIP proc., 1248, 25 
\bibitem[Kargaltsev et al.(2007)]{Karga2007c} Kargaltsev,~O., Pavlov,~G.~G., \& Garmire,~G.~P. 2007, ApJ, 660, 1413 
\bibitem[Kargaltsev et al.(2009)]{Karga2009} Kargaltsev,~O., Pavlov,~G.~G., \& Wong,~J.~A. 2009, ApJ, 690, 891 
\bibitem[Kargaltsev et al.(2005)]{Karga2005} Kargaltsev,~O.~Y., Pavlov,~G.~G., Zavlin,~V.~E., \& Romani,~R.~W. 2005, ApJ, 625, 307 
\bibitem[Kargaltsev et al.(2008)]{Karga2008a} Kargaltsev,~O., Misanovic.~Z., Pavlov,~G.~G., Wong,~J.~A., \& Garmire,~G.~P. 2008, ApJ, 684, 542 
\bibitem[Kennel \& Coroniti(1984)]{Kennel1984} Kennel,~C.~F., \& Coroniti,~F.~V. 1984, ApJ, 283, 710
\bibitem[Komissarov \& Lyubarsky(2004)]{Komissarov2004} Komissarov,~S., \& Lyubarsky,~Y. 2004, Ap\&SS, 293, 107
\bibitem[Lemoine-Goumard et al.(2011)]{Lemoine2011} Lemoine-Goumard,~M., et al. 2011, preprint (arXiv1108.0161)
\bibitem[Linsley(1975)]{Linsley1975} Linsley,~J. 1975, ICRC proc., Vol.2, 592  
\bibitem[Maccacaro et al.(1988)]{Maccacaro1988} Maccacaro,~T., Gioia,~I.~M., Wolter,~A., Zamorani,~G., \& Stocke,~J.~T. 1988, ApJ, 326, 680
\bibitem[Mignani et al.(2011)]{Mignani2011} Mignani, R.\ P., Shearer, A., De Luca, A., Moran, P., Collins, S., \& Marelli, M. 2011, preprint (arXiv:1108.0176)
\bibitem[Mori et al.(2008)]{Mori2008} Mori,~K., Kargaltsev,~O. Pavlov,~G., Yamamoto,~M., Shibata,~S., \& Tsunemi,~H. 2008, in 37th COSPAR Scientific Assembly, Montreal, Canada, p.2105 
\bibitem[Mori et al.(2001)]{Mori2001} Mori,~K., Tsunemi,~H., Miyata,~E., Baluta,~C.~J., Burrows,~D.~N., Garmire,~G.~P., \& Chartas,~G. 2001, in ASP Conf. Ser. 251, New Century of X-Ray Astronomy, ed. H. Inoue \& H. Kunieda (San Francisco:ASP), 576
\bibitem[Pavlov et al.(1999)]{Pavlov1999} Pavlov,~G.~G., Zavlin,~V.~E., \& Tr$\rm \ddot{u}$mper,~J. 1999, ApJ, 511, L45 
\bibitem[Pavlov et al.(2006)]{Pavlov2006} Pavlov,~G.~G., Sanwal,~D. \& Zavlin,~V.~E. 2006, ApJ, 643, 1146 
\bibitem[Pavlov et al.(2008)]{Pavlov2008} Pavlov,~G.~G., Kargaltsev,~O., \& Brisken,~W.~F. 2008, ApJ, 675, 683 
\bibitem[Pavlov et al.(2010)]{Pavlov2010} Pavlov,~G.~G., Bhattacharyya,~S., \& Zavlin,~V.~E. 2010, ApJ, 715, 66 
\bibitem[Pavlov et al.(1994)]{Pavlov1994} Pavlov,~G.~G., Shibanov,~Y,~A., Ventura,~J., \& Zavlin,~V.~E. \ 1994, A\&A, 289, 837 
\bibitem[Pavlov et al.(1995)]{Pavlov1995} Pavlov,~G.~G., Shibanov,~Y,~A., Zavlin,~V.~E. \& Meyer,~R.~D. 1995, in The Lives of the Neutron Stars, ed. M. A. Alpar, U. Kiziloglu, \& J. van Paradijs (NATO ASI Ser. C, 450; Dordrecht:Kluwer), 71 
\bibitem[Pavlov et al.(2003)]{Pavlov2003} Pavlov,~G.~G., Teter,~M.~A., Kargaltsev,~O. \& Sanwal,~D. 2003, ApJ, 591, 1157 
\bibitem[Pavlov et al.(2001)]{Pavlov2001} Pavlov,~G.~G., Zavlin,~V.~E., Sanwal,~D., Burwitz,~V., \& Garmire,~G.~P. 2001, ApJ, 552, L129 
\bibitem[Pellizzoni et al.(2009)]{Pellizzoni2009} Pellizzoni,~A., et al. 2009. ApJ, 695, L115
\bibitem[Renaud et al.(2008)]{Renaud2008} Renaud,~M., Hoppe,~S., Komin,~N., Moulin,~E., Marandon,~V., \& Clapson,~A.-C. 2008, AIP proc., 1085, 285 
\bibitem[Roberts et al.(2007)]{Roberts2007} Roberts,~M.~S.~E., Gotthelf,~E.~V., Halpern,~J.~P., Brogan,~C.~L., \& Ransom,~S.~M. 2007, in Becker,~W., Huang,~H.-H., eds, MPE Report 291, Proc. 363 WE-Heraeus Seminar on Neutron Stars and Pulsars. Max-Planck Institute für extraterrestrische Physik , Garching , p. 24 
\bibitem[Romani et al.(2005)]{Romani2005} Romani,~R.~W., Ng,~C.-Y., Dodson,~R., \& Brisken,~W. 2005, ApJ, 631, 480
\bibitem[Sanwal et al.(2002)]{Sanwal2002} Sanwal,~D., Pavlov,~G.~G., Kargaltsev,~O.~Y., Garmire,~G.~P., Zavlin,~V.~E., Burwitz,~V., Manchester,~R.~N., \& Dodson,~R. 2002, in ASP Conf. Ser. 271, Neutron Stars in Supernova Remnants, ed. P. O. Slane \& B. M. Gaensler (San Francisco:ASP), 353 
\bibitem[Skrutskie et al.(2006)]{Skrutskie2006} Skrutskie,~M.~F., et al. 2006, AJ, 131, 1163 
\bibitem[Tsunemi et al.(2001)]{Tsunemi2001} Tsunemi,~H., Mori,~K., Miyata,~E., Baluta,~C., Burrows,~D.~N., Garmire,~G.~P., \& Chartas,~G. 2001, ApJ, 554, 496 
\bibitem[van der Swaluw(2005)]{Swaluw2005} van der Swaluw,~E. 2005, Advances in Space Research, 35, 1123 
\bibitem[Yakovlev \& Pethick(2004)]{Yakovlev2004} Yakovlev,~D.~G., \& Pethick,~C.~J. 2004, ARA\&A, 42, 169 
\bibitem[Zacharias et al.(2005)]{Zacharias2005} Zacharias,~N., Monet,~D.~G., Levine,~S.~E., Urban,~S.~E., Gaume,~R., \& Wycoff,~G.~L. 2005, VizieR Online Data Catalog, 1297, 0 
\bibitem[Zavlin(2007)]{Zavlin2007} Zavlin,~V.~E. 2007, ApJ, 665, L143 
\bibitem[Zavlin \& Pavlov(2004)]{Zavlin2004} Zavlin,~V.~E., \& Pavlov,~G.~G. 2004, ApJ, 616, 452 
\bibitem[Zhu et al.(2011)]{Zhu2011} Zhu,~W.~W., Kaspi,~V.~M., McLaughlin,~M.~A., Pavlov,~G.~G., Ng,~C.~-Y., Manchester,~R.~N., Gaensler,~B.~M., \& Woods,~P.~M. 2011, ApJ, 734, 44 
\end{thebibliography}
\end{document}